\documentclass[utf8]{FrontiersinHarvard} 

\usepackage{url,hyperref,lineno,microtype,subcaption}
\usepackage[onehalfspacing]{setspace}
\usepackage{float}

\def\keyFont{\fontsize{8}{11}\helveticabold }
\def\firstAuthorLast{Feng {et~al.}} 
\def\Authors{Yanan Feng\,$^{1, 2}$, Xiaohu Li\,$^{1, 3, *}$,Tom J. Millar\,$^{4}$, Ryszard Szczerba\,$^{5}$, Ke Wang\,$^{6}$, Donghui Quan\,$^{7, 1}$, Shengli Qin\,$^{8, 1}$, Xuan Fang\,$^{9, 1, 10, 11}$, Juan Tuo\,$^{1, 2}$, Zhenzhen Miao\,$^{1}$, Rong Ma\,$^{1, 2}$, Fengwei Xu\,$^{6, 12}$, Jingfei Sun\,$^{1}$, Biwei Jiang\,$^{13, 14}$, Qiang Chang\,$^{15}$, Jianchao Yang\,$^{1, 2}$, Gao-Lei Hou\,$^{16}$, Fangfang Li\,$^{1}$, and Yong Zhang\,$^{17, 1,11}$}
\def\Address{
$^{1}$ Xinjiang Astronomical Observatory, Chinese Academy of Sciences, China  \\
$^{2}$ University of Chinese Academy of Sciences, China \\
$^{3}$ Key Laboratory of Radio Astronomy, Chinese Academy of Sciences, China  \\
$^{4}$ Astrophysics Research Centre, School of Mathematics and Physics, Queen's University Belfast, UK  \\
$^{5}$ Nicolaus Copernicus Astronomical Center, Polish Academy of Sciences, Poland    \\
$^{6}$ The Kavli Institute for Astronomy and Astrophysics at Peking University (KIAA-PKU), China   \\
$^{7}$ Research Center for Intelligent Computing Platforms, Zhejiang Laboratory, China  \\
$^{8}$ Department of Astronomy, Yunnan University, China   \\
$^{9}$ National Astronomical Observatories, Chinese Academy of Sciences, China \\
$^{10}$ Department of Physics, Faculty of Science, The University of Hong Kong, China \\
$^{11}$ Laboratory for Space Research, Faculty of Science, The University of Hong Kong, China\\
$^{12}$ Department of Astronomy at Peking University (DoA-PKU), China  \\
$^{13}$ Institute for Frontiers in Astronomy and Astrophysics, Beijing Normal University, China  \\
$^{14}$ Department of Astronomy, Beijing Normal University, China  \\
$^{15}$ School of Physics and Optoelectronic Engineering, Shandong University of Technology, China \\
$^{16}$ MOE Key Laboratory for Non-Equilibrium Synthesis and Modulation of Condensed Matter, School of Physics, Xi'an Jiaotong University, China  \\
$^{17}${ School of Physics and Astronomy, Sun Yat-sen University, China}
}

\def\corrAuthor{Xiaohu Li}

\def\corrEmail{xiaohu.li@xao.ac.cn, 150 Science 1-Street, Urumqi, Xinjiang 830011, China}

\usepackage{soul}

\begin{document}
\onecolumn
\firstpage{1}


\title{ Photochemical origin of SiC$_2$ in the circumstellar envelope of carbon-rich AGB stars revealed by ALMA}  

\author[\firstAuthorLast ]{\Authors} 
\address{} 
\correspondance{} 

\extraAuth{}

\maketitle

\begin{abstract}
\section{}
Whether SiC$_2$ is a parent species, that is formed in the photosphere or as a by-product of high-temperature dust formation, or a daughter species, formed in a chemistry driven by the photodestruction of parent species in the outer envelope, has been debated for a long time.
Here, we analyze the ALMA observations of four SiC$_2$ transitions in the CSEs of three C-rich AGB stars (AI Vol, II Lup, and RAFGL 4211), and found that SiC$_2$ exhibits an annular, shell-like distribution in these targets, suggesting that SiC$_2$ can be a daughter species in the CSEs of carbon-rich AGB stars. 
The results can provide important references for future chemical models.

\tiny
 \keyFont{ \section{Keywords:} AGB stars, circumstellar envelope(CSEs), daughter species, mass loss, column density} 
\end{abstract}

\section{Introduction}

Late evolutionary stars with masses between 0.8 and 8 $ M\rm_{\odot}$ \citep{hofner2018mass} experience dramatic mass losses. The stellar wind continues to eject material outward, eventually forming a circumstellar envelope (CSE), often referred to as the molecular space factory. According to different C/O ratios, asymptotic giant branch (AGB) stars are divided into three classes: O-rich AGB stars (oxygen-rich, C/O ${<}$ 1), S-type AGB stars (C/O ${\approx}$ 1), and C-rich AGB stars (carbon-rich, C/O ${>}$ 1). 
Over 105 molecular species have been detected in the circumstellar envelope of the evolved stars \citep{Decin+2021}. These species, distributed in different areas of the CSEs, can help us to trace shell properties, such as local temperature and gas composition, and help us to understand the chemical synthesis within the objects. A large number of carbon-bearing species are observed to be abundant surrounding the carbon stars, for instance, CO, CS, HC$_3$N, and C$_2$H, some of which come from the inner layers of the CSEs, and some are formed in the outer regions \citep{2008A&A...480..431D, 2016A&A...588A...4L}.

The silacyclopropynylidene (SiC$_2$) was first detected and confirmed in IRC+10216 by \citet{1984ApJ...283L..45T} who derived a fairly large column density, 1.5 $\times$ 10$^{14}$ cm$^{-2}$. \citet{1986A&A...157...35G} suggested that SiC$_2$ was formed by the reaction of Si$^+$ with C$_2$H$_2$ or C$_2$H followed by dissociative recombination with electrons. \citet{glassgold1991formation} suggested alternative pathways, initiated by reactions of SiS with C$_2$H$_2^+$ or C$_2$H$_3^+$. 
All these routes need the presence of UV photons to form cations and radicals, and thus treat SiC$_2$ as a daughter species.

The spatial distribution of SiC$_2$ in IRC+10216 was discussed by \citet{1992PASJ...44..469T} and \citet{1993A&A...280L..19G} who showed that it showed a spherical shell-like structure indicative of a daughter species. \citet{1995Ap&SS.224..293L} and \citet{Gensheimer95} used the Plateau de Bure and Berkeley Illinois Maryland Array (BIMA), respectively, to map SiC$_2$ emission, showing that it had a shell structure. The first confirmation that SiC$_2$ was present close to the star came from \citet{cernicharo2010high} who used the Herschel Space Observatory to detect some 55 transitions from energy levels 500-900 K above ground state, implying that SiC$_2$ was present in the dust-forming zone. Subsequently \citet{2014MNRAS.445.3289F}, using the CARMA interferometer, and \citet{prieto2015si} using Atacama Large Millimeter Array (ALMA) detected emission from both the central region and the outer shell.

In a work devoted to single-dish observations of 25 C-rich AGB stars \citep{2018A&A...611A..29M}, SiC$_2$ and SiC were detected in about half of the sources. They found that the abundance of SiC$_2$ decreased with increasing envelope density
increasing envelope density, indicating that SiC$_2$ is a parent species in these CSEs. 
This finding confirms SiC$_2$ to be a parent species in CSE. \citet{2020A&A...642A..20D} detected SiC$_2$ in S-type AGB star W Aql, and they showed that the emission occurs in a ring of radius 1-2$^{\prime \prime}$.

Most evidence shows that SiC$_2$ is formed in LTE conditions close to the star, although IRC+10216 has an increase in abundance of SiC$_2$ at around 10$^{\prime \prime}$ indicating an additional formation in the outer regions of CSE. To date, no millimeter-wave interferometric array observation of SiC$_2$ have been published in CSE of C-rich AGB stars other than IRC+10216. In this work, we provide evidence that SiC$_2$ seems to be a daughter molecule through high-resolution observations of three C-rich AGB stars. We describe the observations in Sect.\ref{sec:part2}. The results are discussed in Sect.\ref{sec:part3}. The conclusions are presented in Sect.\ref{sec:part4}. The spatial distribution and the fractional abundance of SiC$_2$ in AI Vol together with $5_{0,\,5}-4_{0,\,4}$ transition in RAFGL 4211 are discussed in the main paper, while the rest of the results are presented in the Supplementary Material.

\section{Observations} \label{sec:part2}

Observations were made using an ALMA 12 m array. The spectral line observations in Band 3 of AI Vol, RAFGL 4211, and II Lup covering four Windows with a bandwidth of 2 GHz were performed on 2015 August 16 (2013.1.00070.S, PI: Nyman, Lars-$\rm \AA$ke). The data was extracted from ALMA Archive\footnote {\url  {https://almascience.nao.ac.jp/aq}}. For AI Vol, the configuration used for observation is a 12 m main array, baselines: 21 - 783 m. The image cube per spw averages every 2 $\times$ 488 kHz channels. The rms requirement is 3 mJy per 1 arcsec beam, per 0.9 MHz ($\sim$ 2.7 km s$^{-1}$). The synthesized rms is 1.3 mJy/beam, and the synthesized beam is 1.4$^{\prime \prime}$ $\times$ 0.9$^{\prime \prime}$. The parameters of the observed transitions are collected in Table \ref{table:2}. The frequency resolution during observation is 1128.9984 kHz, corresponding to the velocity resolution of 2.526 $-$ 3.122 km s$^{-1}$. The wide frequency range covers four lines of SiC$_2$ ($4_{0,\,4}-3_{0,\,3}$, $4_{2,\,3}-3_{2,\,2}$, $4_{2,\,2}-3_{2,\,1}$, and $5_{0,\,5}-4_{0,\,4}$).

AI Vol, RAFGL 4211, and II Lup are all carbon-rich AGB stars with mass loss. 
A survey conducted by \citet{smith2015molecular} at the 3 mm band revealed that the three carbon-rich stars possess abundant molecules, and the sensitive detection of molecular spatial distributions relies on more advanced telescopes such as ALMA. 
In this study, we present the physical parameters taken from the literature and the synthesized beam for the observed sources, which are listed in Table \ref{table:1}.

The imaging process of the calibrated data was performed using CASA 4.3.1\footnote{\url{https://casa.nrao.edu/}} software manually. The ``clean" task was employed, and to achieve a balance between spatial resolution and noise gain in the resulting image, the Briggs weighting function was applied along with adjustments to the ``robust" parameter of 0.5. The rms was calculated near the center of the field, excluding all emissions within the region. The rms values for all channel maps were obtained using the task ``imstat'' in CASA software. Each pixel on the image plane corresponded to a size of 0.2 arcseconds. Then we used Python software to map and analyze ALMA product data. To calculate the molecular column density, we employed the shell fitting in the GILDAS software\footnote{\url{http://www.iram.fr/IRAMFR/GILDAS}}, along with the splatalogue 
 \footnote{\url{https://splatalogue.online//advanced.php}} and the CDMS databases\footnote{\url{http://www.astro.uni-koeln.de/cdms/catalog}} \citep{Muller+2005}.

\section{Result} \label{sec:part3}

\subsection{Spatial distribution of SiC$_2$} \label{sec:part3.1}

We obtained four transitions of SiC$_2$ in each source, except for the $4_{2,\,2}-3_{2,\,1}$ transition, which was not observed in II Lup. Based on the $S\mu^2$ from Table \ref{table:4}, the $4_{2,\,2}-3_{2,\,1}$ transition of SiC$_2$ in the 3 mm wavelength range ranked third in terms of signal intensity. However, we did not detect any signal. Our data were obtained from the ALMA Archive, and from the data information, we found that the other three transitions of SiC$_2$ and the data from the two additional sources had map sizes of 800 $\times$ 800 or 640 $\times$ 640 pixels. In contrast, the data without signal detection only covered a map size of 300 $\times$ 300 pixels. We speculate that there may have been some unknown special circumstances during the observations that led to the absence of signal.

\begin{figure}[h!]
\begin{center}
\includegraphics[width=18cm]{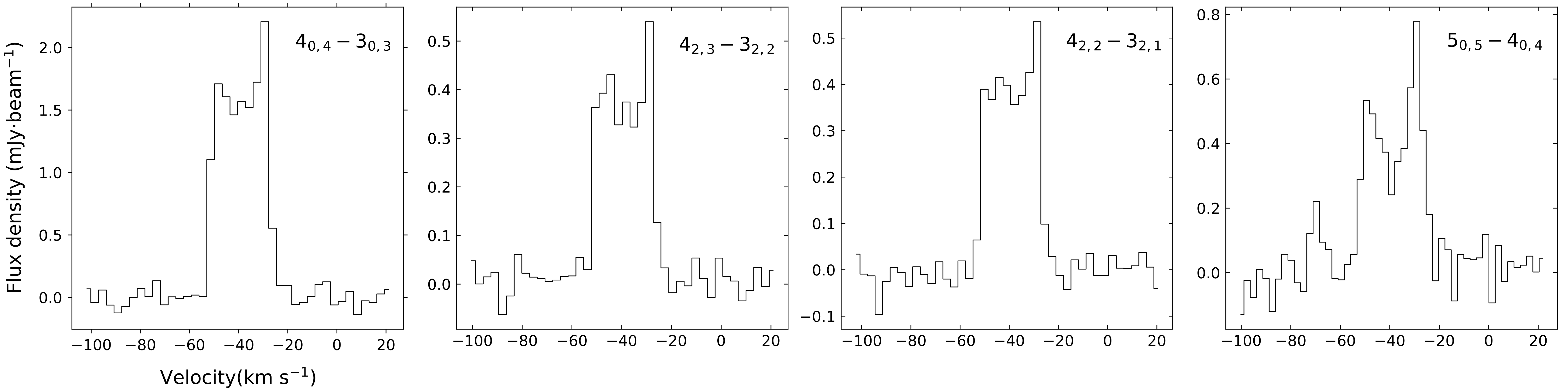}
\end{center}
\caption{The spectral line shapes of SiC$_2$ toward AI Vol, 
integrated with a 100 $\times$ 100 pixel ranges centered on the star.}
\label{fig:6}
\end{figure}

AI Vol does not have the strongest spectral line signal, but the signal-to-noise ratio of spatial brightness distribution is higher than that of the other two stars. \citet{2015A&A...575A..91C} have reported the observations and model results of a C-rich AGB star, IRC+10216. They suggest that a companion star may explain the spiral structure of its CSE. \citet{2018MNRAS.480.1006L} reported spiral structures of several parent molecules in II Lup, suggesting that the companion star is the main forming mechanism of mass loss. The results show that the spectral line signal of AI Vol is stronger than RAFGL 4211 (in Figures \ref{fig:6} and \ref{subfig:9}). In order to get reliable results, we concentrate on analyzing the distribution around the center star of AI Vol (in Sect. \ref{sec:part3.1.1}). The rotational diagrams method was used to calculate fractional abundance, and the detailed analysis is presented in Sect. \ref{sec:part3.2}.

\subsubsection{AI Vol}\label{sec:part3.1.1}

\begin{figure}[h!]
\begin{center}
\includegraphics[width=15cm]{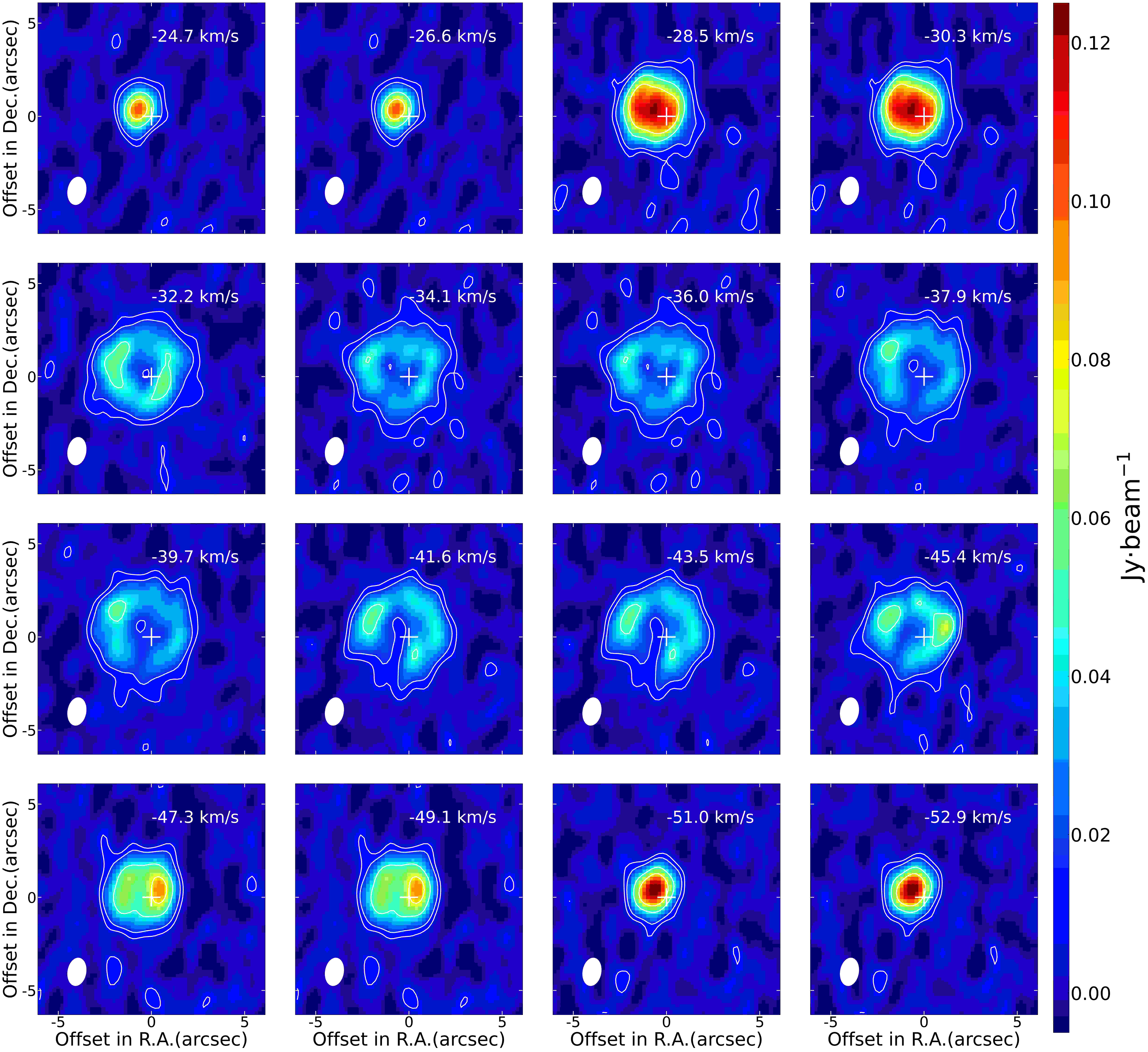}
\end{center}
\caption{Channel maps of SiC$_2$ $(4_{0,\,4} - 3_{0,\,3})$ towards AI Vol. The shape of the synthesized beam is shown in the lower-left corner of each panel with a size of 1.51 $\times$ 0.99 $^{\prime \prime}$, with PA 9.82$^{\circ}$. The systemic velocity is displayed in the top-right corner of each panel. The white cross represents the position in the map. The white contour maps display the flux levels of SiC$_2$ transition at 5, 10, 30, and 50 times the rms noise, 1$\sigma$ = 1.6 mJy$\cdot$beam$^{-1}$. The selected range of color-bar can show the spatial distribution of molecules better. The brightness distribution of the elongated cavity is similar to that in IRC+10216 \citep{1995Ap&SS.224..293L}.}
\label{fig:1}
\end{figure}

\begin{figure}[h!]
\begin{center}
\includegraphics[width=15cm]{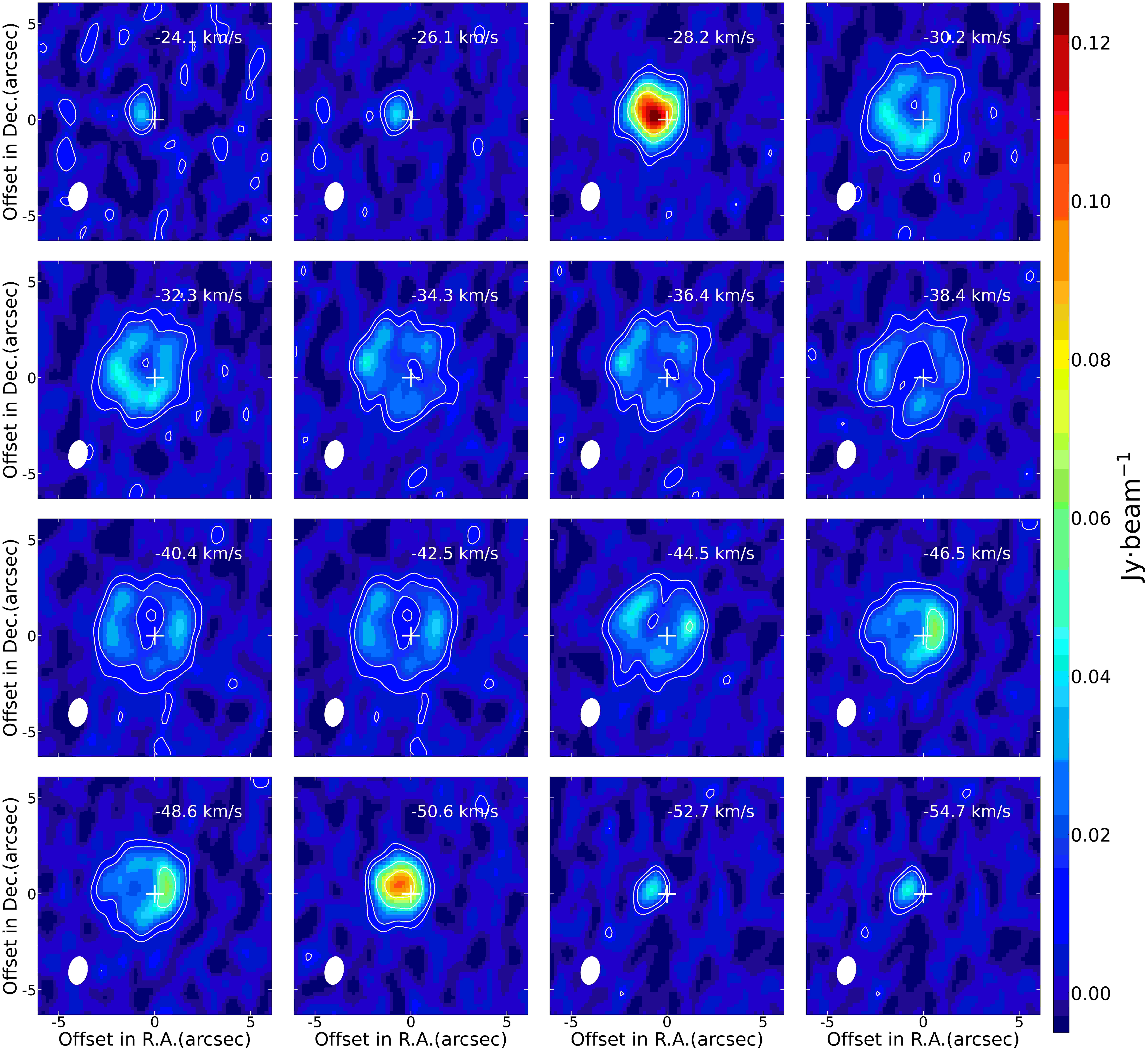}
\end{center}
\caption{Channel maps of SiC$_2$ $(4_{2,\,2} - 3_{2,\,1})$ towards AI Vol. The shape of the synthesized beam is shown in the lower-left corner of each panel with a size of 1.48 $\times$ 0.98 $^{\prime \prime}$, with PA 10.44$^{\circ}$. The velocity (km s$^{-1}$) is displayed in the top-right corner of each panel. The white cross represents the center of the star in the map. The white contours display the flux levels of SiC$_2$ transition at 5, 10, 30, and 50 times of the rms noise, 1$\sigma$ = 1.5 mJy$\cdot$beam$^{-1}$.}
\label{fig:2}
\end{figure}

Figures \ref{fig:1} and \ref{fig:2} shows SiC$_2$ ($4_{0,\,4}-3_{0,\,3}$ and $4_{2,\,2} - 3_{2,\,1}$) radial velocity channel map toward C-rich AGB star AI Vol (The other transitions are shown in Figures (\ref{subfig:1} and \ref{subfig:2})). Around the local standard of rest velocity $V\rm_{LSR}$ of AI Vol (-39 km s$^{-1}$), SiC$_2$ shows a ring distribution around the center star with a diameter of $\thicksim$ 4-6$^{\prime \prime}$ ($\thicksim$ 2840.95 $-$ 4258.08 au), and the signal strength reaches 20.4 and 13.7$\sigma$ for $4_{0,\,4}-3_{0,\,3}$ and $4_{2,\,2} - 3_{2,\,1}$ transitions. Based on the images presented by \citet{2015A&A...574A...5D}, it is evident that the parent molecules, such as SiO, exhibit a concentrated distribution within a compact structure surrounding the central star. In contrast, our observation reveals a characteristic pattern specific to the daughter molecules, 
distributed in hollow rings around the star \citep{2015ApJ...814..143A, 2017A&A...601A...4A}. The brightness distributions of Figures \ref{fig:1} and \ref{fig:2} under different velocity components are all coming from the transitions ($4_{0,\,4}-3_{0,\,3}$ and $4_{2,\,2} - 3_{2,\,1}$), without spatial extension caused by the fine structure.

\begin{figure}[h!]
\begin{center}
\includegraphics[width=7.1cm]{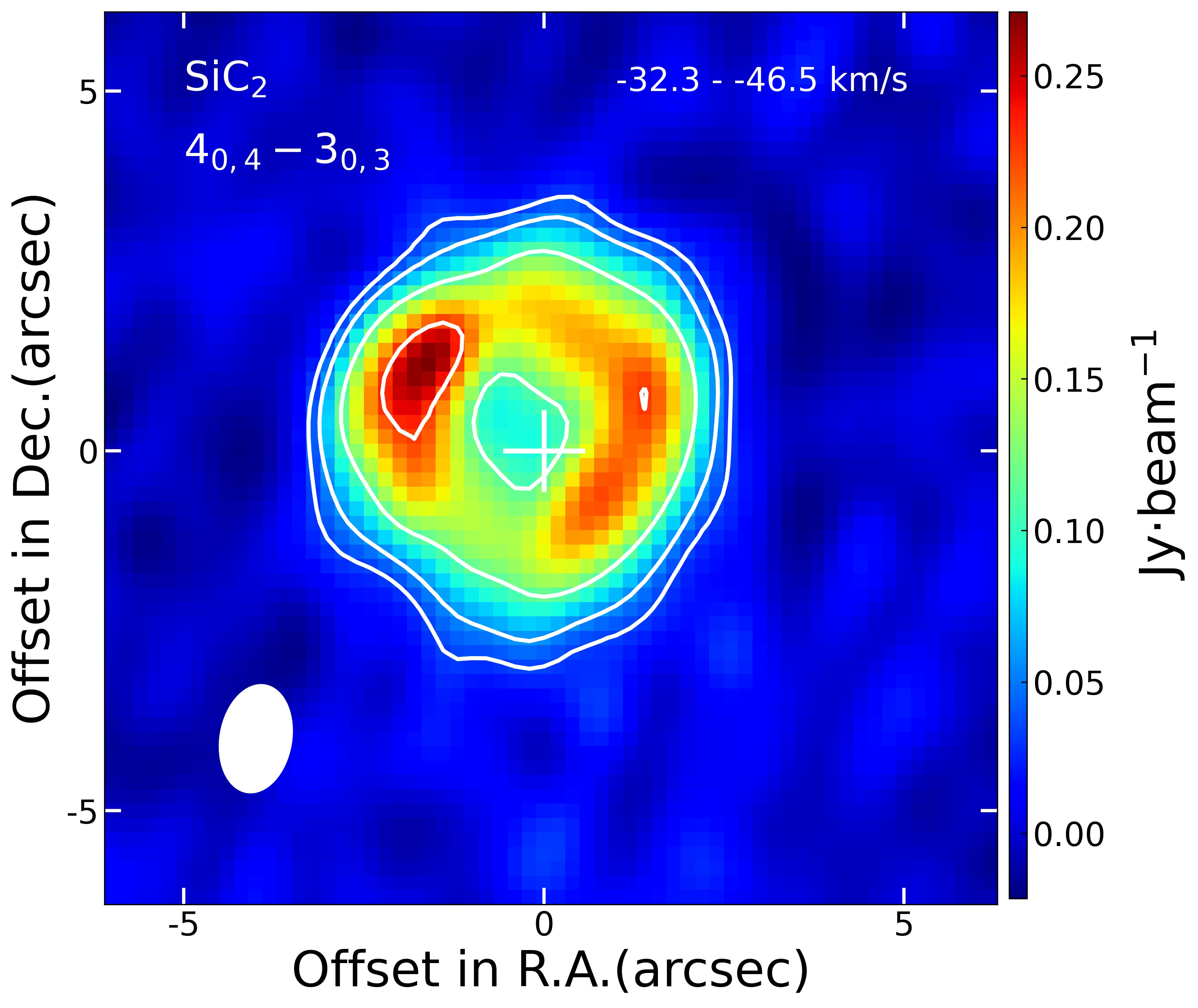}
\includegraphics[width=7.1cm]{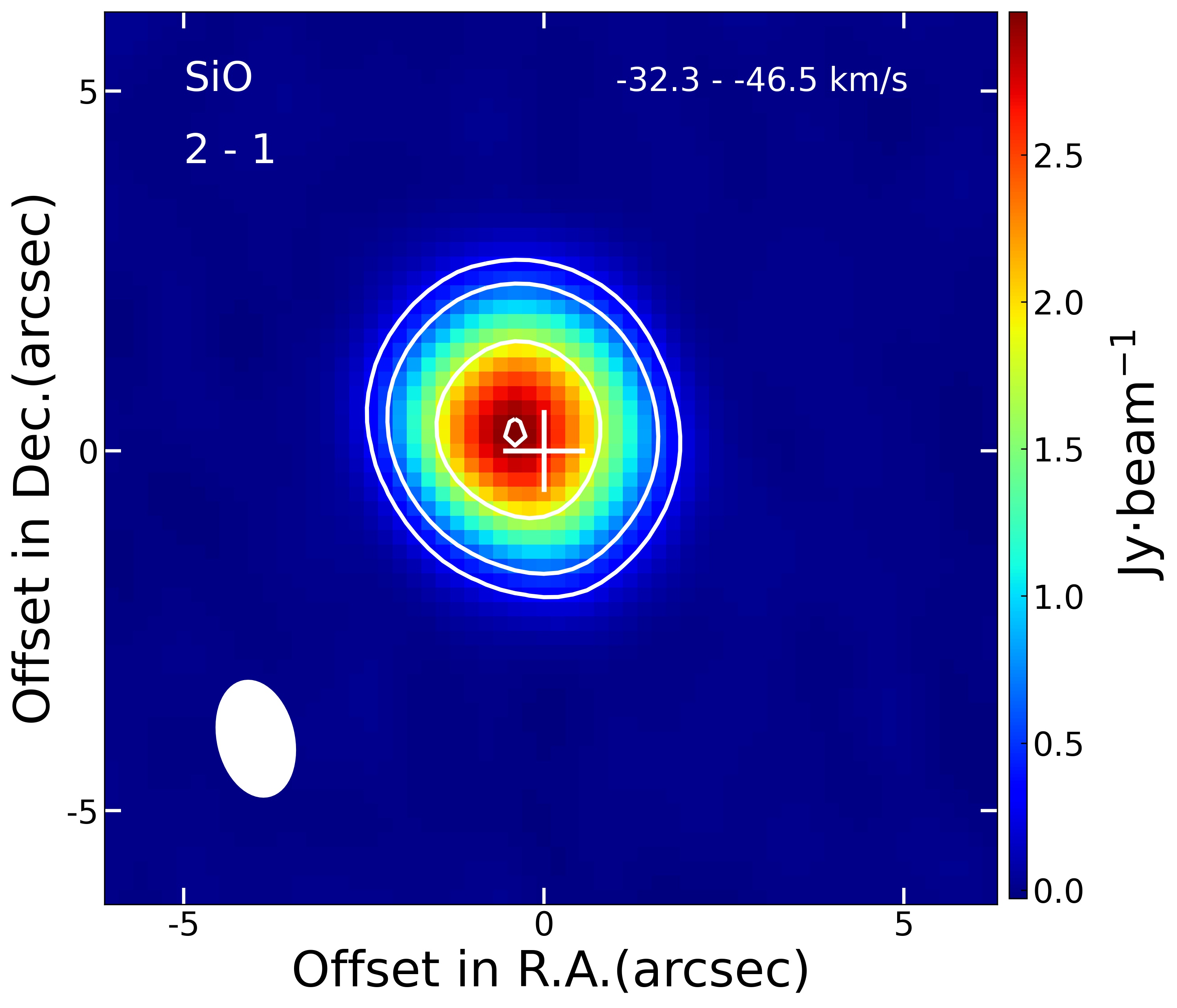}
\end{center}
\caption{Brightness distribution of SiC$_2$ (left panel) and SiO (right panel) transitions at the source velocity (-32.3 $-$ -46.5 km s$^{-1}$) in AI Vol. Color bar shows the brightness distribution of different regions. The white cross represents the center of the star in the map. The beam size of the observation in SiC$_2$ is 1.51 $\times$ 0.99\,$^{\prime \prime}$, with PA 9.82$^{\circ}$. 
The beam size of the observation in SiO is 1.64 $\times$ 1.06\,$^{\prime \prime}$, with PA -12.99$^{\circ}$. 
The white contours display the flux levels of SiC$_2$ at 3, 5, 10, and 20 times of the rms noise, with 1$\sigma$ = 11.3\,mJy$\cdot$beam$^{-1}$. 
For SiO (2 - 1), the flux levels are at 5, 10, 30, and 50 times of the rms noise, where 1$\sigma$ = 58.9\,mJy$\cdot$beam$^{-1}$.}
\label{fig:3}
\end{figure}

Brightness distribution of the $4_{0,\,4} -3_ {0,\,3}$ transition of SiC$_2$ and the 2 - 1 transition of SiO around $V\rm_{LSR}$ channels (from - 32.3 to - 46.5 km s$^{-1}$) is shown in Figure \ref{fig:3}. We can see that the maximum emission of SiC$_2$ occurs at the radius $\sim$ 3$^{\prime \prime}$, with a hole in the distance from the star $\sim$ 2$^{\prime \prime}$. The studies conducted by \citet{1992PASJ...44..469T} and \citep{2015ApJ...814..143A} showcase the spatial distribution of the daughter species centered around $V\rm_{LSR}$. SiO is present as the parent species in three types of AGB stars \citep{Cherchneff+2006, Ramstedt+etal+2009}. The SiC$_2$ and SiO shown in Figure \ref{fig:3} represent the spatial distribution characteristics of the daughter species and parent species, respectively, exhibiting distinct hollow and compact structures.

\begin{figure}[h!]
\begin{center}
\includegraphics[width=7.1cm]{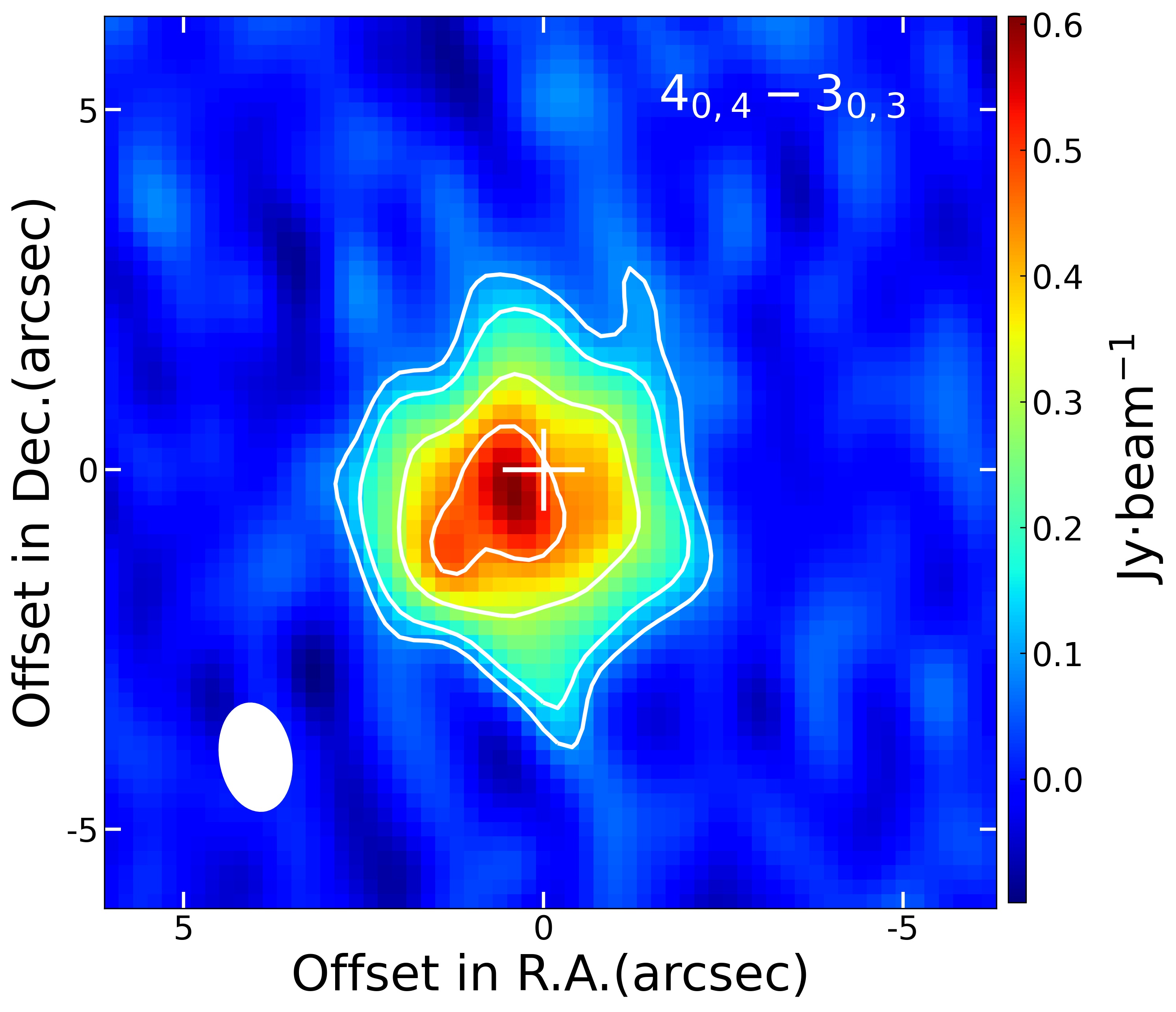}
\includegraphics[width=7.1cm]{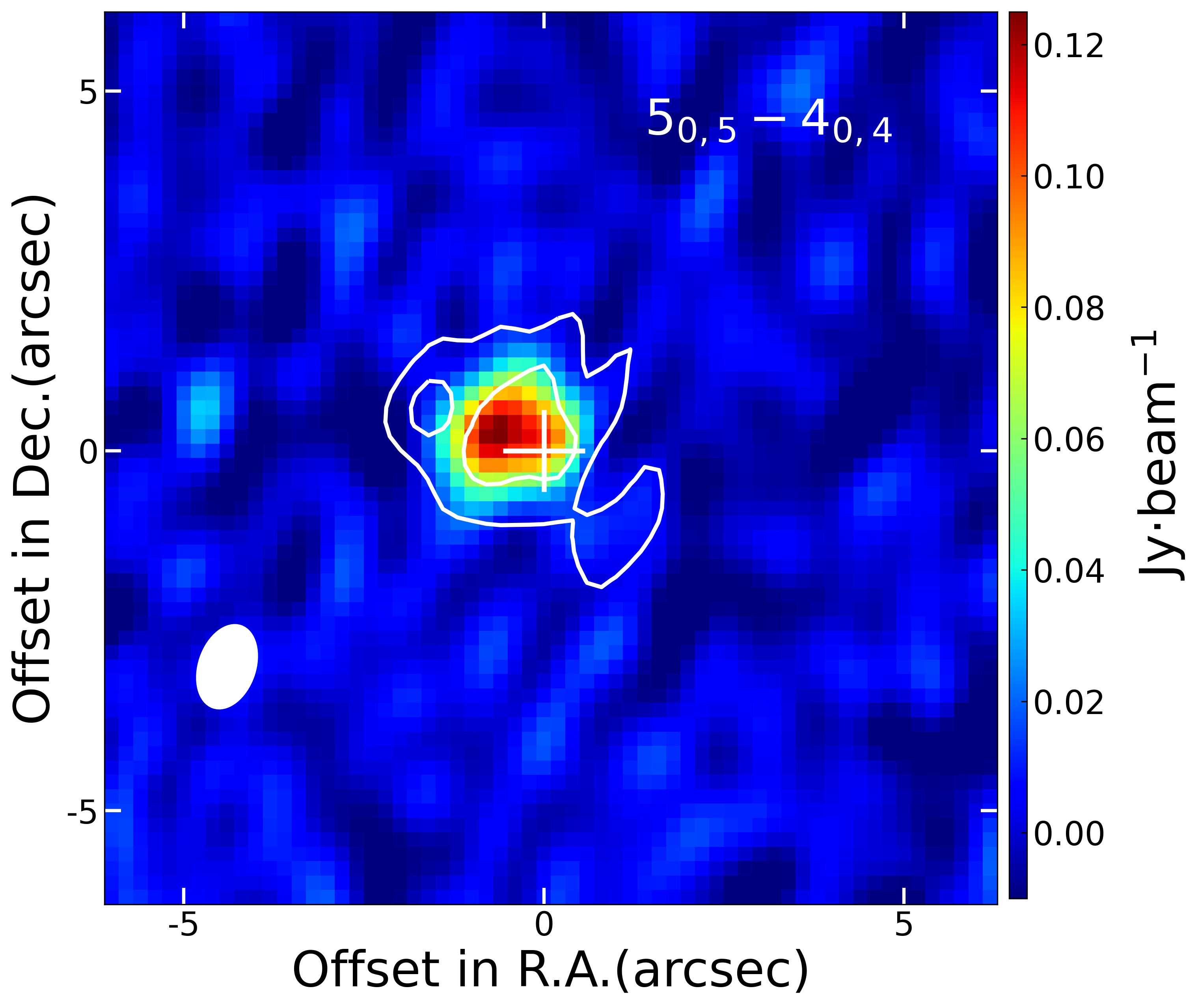}
\end{center}
\caption{The zeroth moment map of the SiC$_2$ $(4_{0,\,4} - 3_{0,\,3}$ and $5_{0,\,5} - 4_{0,\,4})$ line at 93.063639 and 115.382375 GHz towards AI Vol. The beam size of observation is 1.51 $\times$ 0.99$^{\prime \prime}$, with PA 9.82$^{\circ}$. 
The molecular transitions are displayed in the top-right corner of each panel. 
In right panel, the beam size of observation is 1.20 $\times$ 0.79$^{\prime \prime}$, with PA 18.86$^{\circ}$. 
In left panel, the white contours display the flux levels at 3, 5, 10, and 15 times of the rms noise, with 1$\sigma$ = 30.5 mJy$\cdot$beam$^{-1}$. 
In right panel, the flux levels are shown at 3 and 5$\sigma$ (1$\sigma$ = 0.1 Jy$\cdot$beam$^{-1}$).}
\label{fig:4}
\end{figure}

The northeast part of the shell emission has relatively stronger spectral signals. The SiC$_2$ ($4_{0,\,4}-3_{0,\,3}$) transition for AI Vol is stronger than the other transitions, and the hollow shell structure can be seen more clearly. The SiC$_2$ at different velocity components showed an elongated cavity extending from the inner region to a location with a southern radius of $\sim$ 2$^{\prime \prime}$. This structure may be caused by the companion star, but the AI Vol has no spiral structure typical of the companion star \citep{2015A&A...575A..91C}. The gas expands at a lower velocity near the star, and lower-excited state transitions farther from the star are more spatially distributed than higher-excited state transitions in the same source. This is seen in Figure \ref{fig:4} where the zeroth-order moment map of the of the $(5_{0,\,5} - 4_{0,\,4})$ transition has a smaller extent than that of the $(4_{0,\,4} - 3_{0,\,3})$ transition.

The difference between Figure \ref{fig:3} and Figure \ref{fig:4} lies in their representation of the data. In Figure \ref{fig:3}, we present channel maps which provide images of vertical slices through the expanding envelope while Figure \ref{fig:4} shows the line intensity integrated over all velocities. From Figure \ref{fig:3}, we can see the clear feature of SiC$_2$ as a daughter molecule. Figure \ref{fig:4} offers a broader perspective by showcasing the distribution of all signals. More over, it reveals that the gas intensity spatial distribution of AI Vol for different transitions varies in size. Compared to transitions with low rotational quantum numbers, transitions with high rotational quantum numbers are distributed closer to the star.

The inner region of the AGB stars is at $\lesssim$ 20$R_\ast$ (3.89 $\times$ 10$^{14}$ cm), and the intermediate area of the CSE of AI Vol is at $\lesssim$ 70$R_\ast$ (1.36 $\times$ 10$^{15}$ cm) \citep{2008A&A...480..431D}. So the radius of the SiC$_2$ molecules in the AI Vol coincides with the region of the daughter species in the model \citep{2016A&A...588A...4L}.

\subsubsection{RAFGL 4211 and II Lup}\label{sec:part3.1.2}

\begin{figure}[h!]
\begin{center}
\includegraphics[width=15cm]{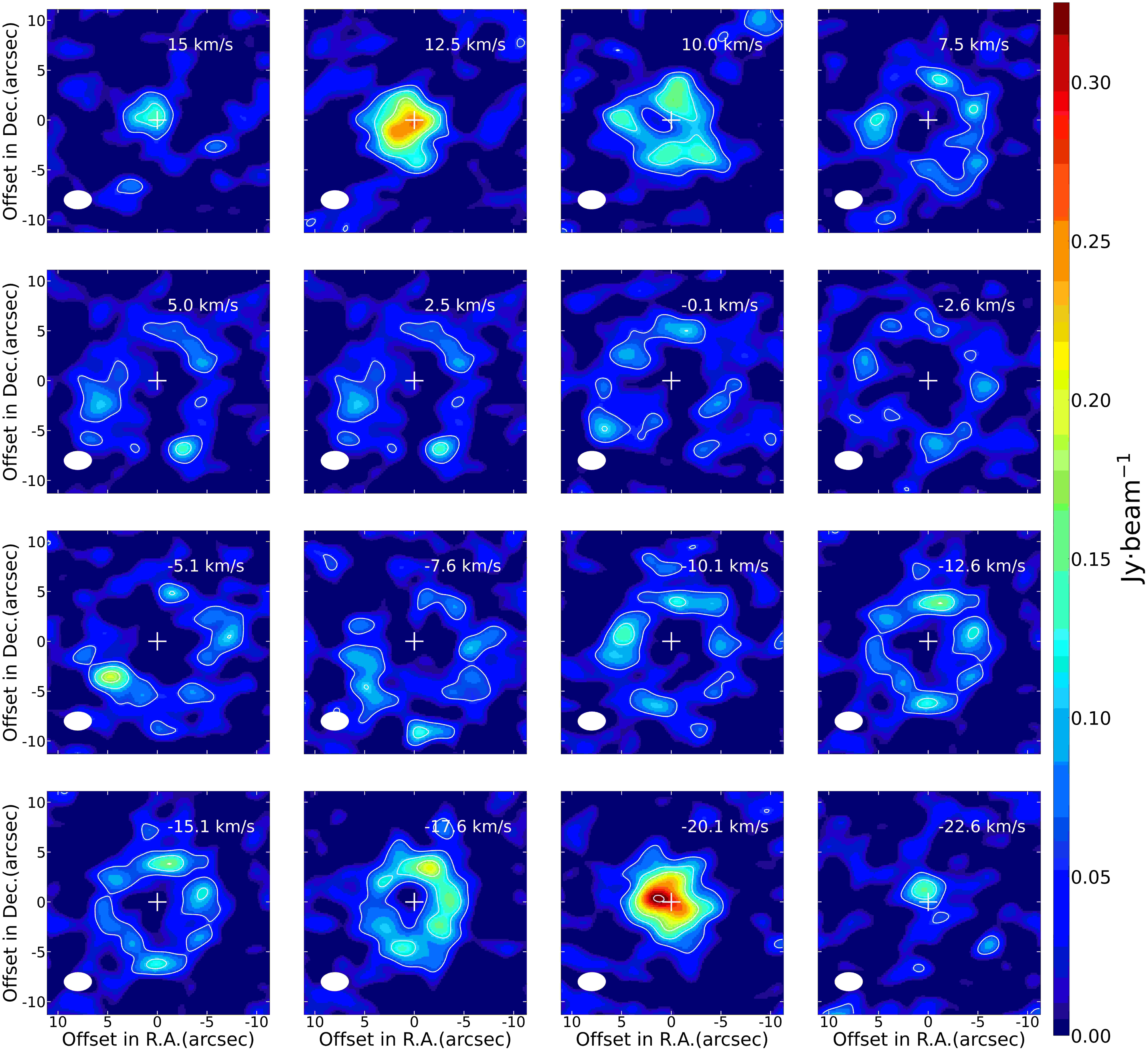}
\end{center}
\caption{Channel maps of SiC$_2$ $(5_{0,\,5} - 4_{0,\,4})$ towards RAFGL 4211. The shape of the synthesized beam is shown in the lower-left corner of each panel with a size of 1.40 $\times$ 0.93 $^{\prime \prime}$, with PA 89.13$^{\circ}$. The white cross represents the center of the star in the map. The white contours delineate the flux levels of the SiC$_2$, indicating increments of 5, 10, 15, and 30 times the rms noise (1$\sigma$ = 11.0 mJy$\cdot$beam$^{-1}$).}
\label{fig:5}
\end{figure}

Figure \ref{fig:5} and Figures (\ref{subfig:3}, \ref{subfig:4}, \ref{subfig:5}) show the channel maps of the four transitions of SiC$_2$ for RAFGL 4211. The synthesized beam of observation is 1.00 $\times$ 1.00 $^{\prime \prime}$, with position angles (PA) 0$^{\circ}$. The signal strengths range from 5$\sigma$ to 8$\sigma$. Brightness distribution radius is $\sim$5$^{\prime \prime}$ around $V\rm _{LSR}$ = - 3.0 km s$^{-1}$ velocity (distance from star center to peak intensity). A hollow shell structure exists within the radius $\sim$3$^{\prime \prime}$ of the star. The SiC$_2$ spatial distribution range of RAFGL 4211 is found to be $\sim$6.36 $\times$ 10$^{16}$ - 8.90 $\times$ 10$^{16}$ cm. In addition to focusing on the brightest component, we see some clump distributions in the outer regions.

Figures (\ref{subfig:6}, \ref{subfig:7}, \ref{subfig:8}) show the channel map of the three transitions of II Lup. The emission $4_{2,\,2} - 3_{2,\,1}$ has no obvious signal and linewidth (the third panel at the bottom of Figure \ref{subfig:9}). The synthesized beam of observation is 0.75 $\times$ 0.45 $^{\prime \prime}$. The signal strengths range from 6$\sigma$ to 11$\sigma$. Figure \ref{subfig:8} shows that $5_{0,\,5}-4_{0,\,4}$ emission has the form of a ring around the central position. The brightness distribution radius of the II Lup near $V\rm_{LSR}$ = 15.5 km s$^{-1}$ velocity is $\sim$10$^{\prime \prime}$. The SiC$_2$ spatial distribution range of II Lup is found to be $\sim$3.74 $\times$ 10$^{16}$ - 7.48 $\times$ 10$^{16}$ cm. As with RAFGL 4211, some clumps can be seen at $\sim$ 16$^{\prime \prime}$. This clump may be part of a spiral structure caused by a companion star.

\subsubsection{Summary of morphology}\label{sec:part3.1.3}

The SiC$_2$ channel map (Figure \ref{fig:2} and \ref{fig:5}) of AI Vol and RAFGL 4211 shows distinct daughter species characteristics. The SiC$_2$ radius greater than 20$R_\ast$ distribution in the three C-rich AGB stars is consistent with the distribution of daughter species in O-rich AGB stars \citep{li2014photodissociation}. The distribution in hollow rings of the three sources indicates that SiC$_2$ is formed in the star's outer layer through chemical reactions. This is the same as the ring distribution of SiC$_2$ in IRC+10216 \citep{1992PASJ...44..469T}. \citet{Gensheimer95} reported that the inner and outer radii of SiC$_2$ of IRC+10216 are mainly distributed in the range of $\sim$2$\times$ 10$^{16}$ - 6 $\times$ 10$^{16}$ cm. In the analyzed observations of three sources, the SiC$_2$ spatial distribution range is about 2.12 $\times$ 10$^{16}$ - 8.90 $\times$ 10$^{16}$ cm. To sum up, combining the ring brightness distribution of SiC$_2$ obtained in this work, we confirm that SiC$_2$ is the daughter molecule in the CSEs of C-rich AGB stars.

\subsection{Abundance} \label{sec:part3.2}

We use the rotational diagrams method to estimate the abundance of SiC$_2$ and the spectral line profiles of the four transitions are shown in Figure \ref{fig:6}. Using the parameters in Table \ref{table:2}, the molecular excitation temperature and column density are calculated under the assumption of LTE, and the equation is as follows \citep[e.g.]{zhang2009molecular, wangke2014},

\begin{equation}\label{equ:1}
\ln(\frac{3kW}{8\pi^3{\nu}S\mu^2}) = \ln(\frac{N}{Q})-\frac{E\rm_u}{kT\rm_{ex}},
\end{equation}

Here, $N$ is the total column density of the molecule, $Q$ partition function, $E\rm_u$ upper-level energy, $T\rm_{ex}$ is the excitation temperature, $k$ the Boltzmann constant, $\rm \upsilon(Hz)$ the rest frequency, $W$($\int T\rm_{R}d\upsilon$) spectral line integral intensity, $S\mu^2$ the  
product of the line strength and the square of the electric dipole moment. 
The values of $S \mu^2$ and $E\rm_u/k$ are taken from the splatalogue databases\footnote{\url{https://splatalogue.online//advanced.php}}. As the upper energy levels are close in temperature, the errors on the derived quantities are rather large.

In order to determine the relative abundance of SiC$_2$ to H$_2$, it is necessary to evaluate the average H$_2$ column density within the radius occupied by SiC$_2$. This can be achieved by employing the following equation \citep{gong20151},
\begin{equation}\label{equ:2}
N_{\rm H_2}=\frac{\dot{M}R/V\rm_{exp}}{\pi R^2m\rm_{H_2}}=\frac{\dot{M}}{\pi RV\rm_{exp}\mu m\rm_H}.
\end{equation}

For AI Vol, $\dot{M}$ = 4.9 $\times$ $10^{-6}$ $M\rm_{\odot}$ $\rm yr^{-1}$ \citep{danilovich2018sulphur}, R is the radius of the peak of the highest brightness point in Figure \ref{fig:4}, which is  3$^{\prime \prime}$; $V\rm_{exp}$ the expansion velocity of 12 km s$^{-1}$ \citep{nyman1995molecular}, $m\rm_H$ the mass of hydrogen, $\mu$ the mean molecular weight of 2.8 of \citet{gong20151}. The derived column density of $\rm H_2$ is 5.52 $\times$ $10^{20}$ $\rm cm^{-2}$. In the rotational analysis using Equation \ref{equ:1}, we added error propagating and obtained the error of excitation temperature ($\delta$$T \rm _{ex}$) and column density ($\delta$$N$), with the integral intensity of 1$\sigma$-error. The formulas for $\delta$$T\rm _{ex}$ and $\delta$$N$ are as follows,

\begin{equation}\label{equ:3}
\delta T_{ex}=\frac{1}{m^{2}_R}\delta m_R,
\end{equation}

and

\begin{equation}\label{equ:4}
\delta N=\sqrt{\delta^2_{N,Q}+\delta^2_{N_{C_R}}}.
\end{equation}

The slope, $m\rm_R$, and intercept, $C$$\rm_R$. $\delta$$N$ is the uncertainty contribution from the rotation partition function and the intercept of the rotational diagram.

\begin{figure}[h!]
\begin{center}
\includegraphics[width=6cm]{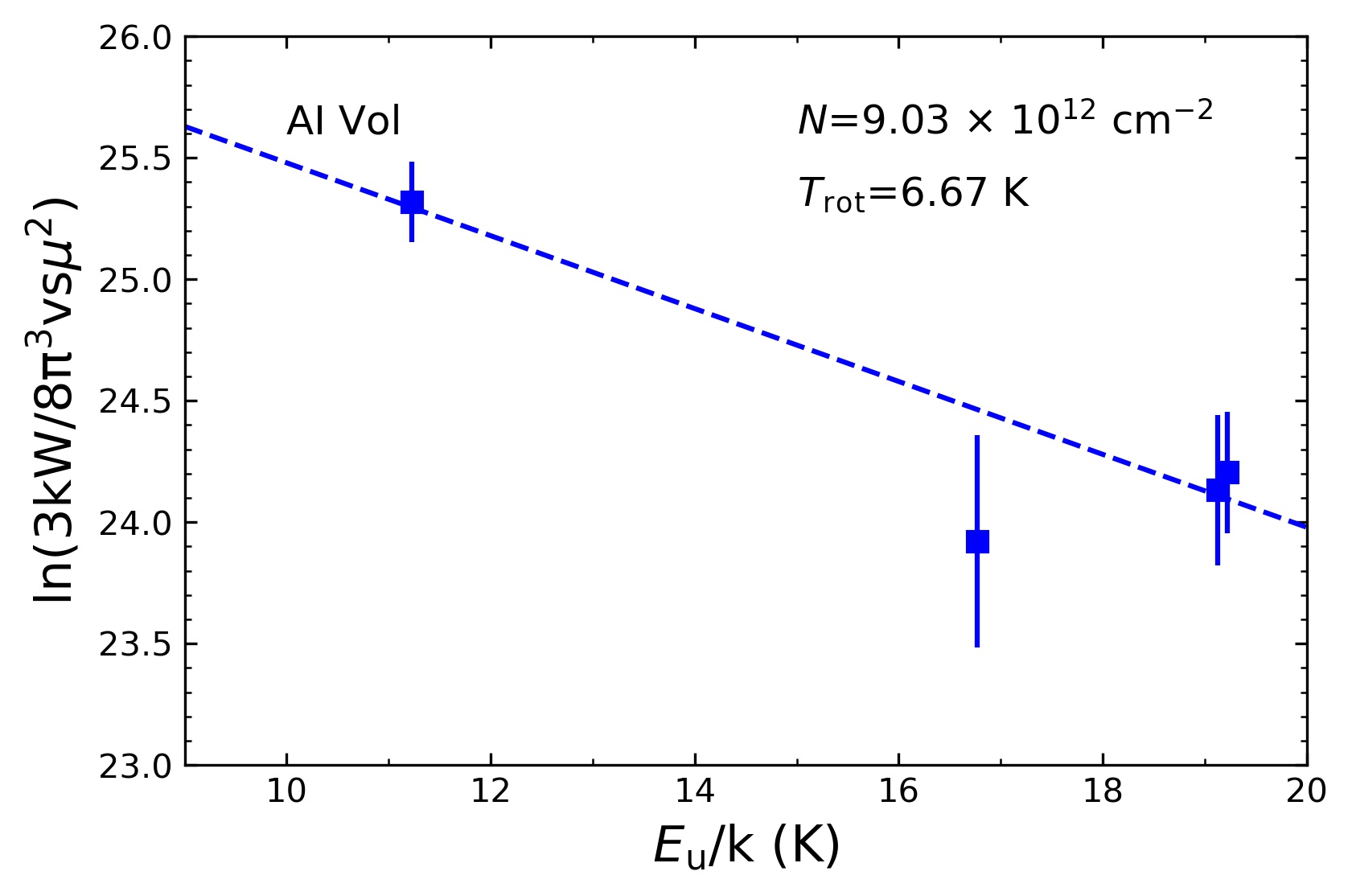}
\includegraphics[width=6cm]{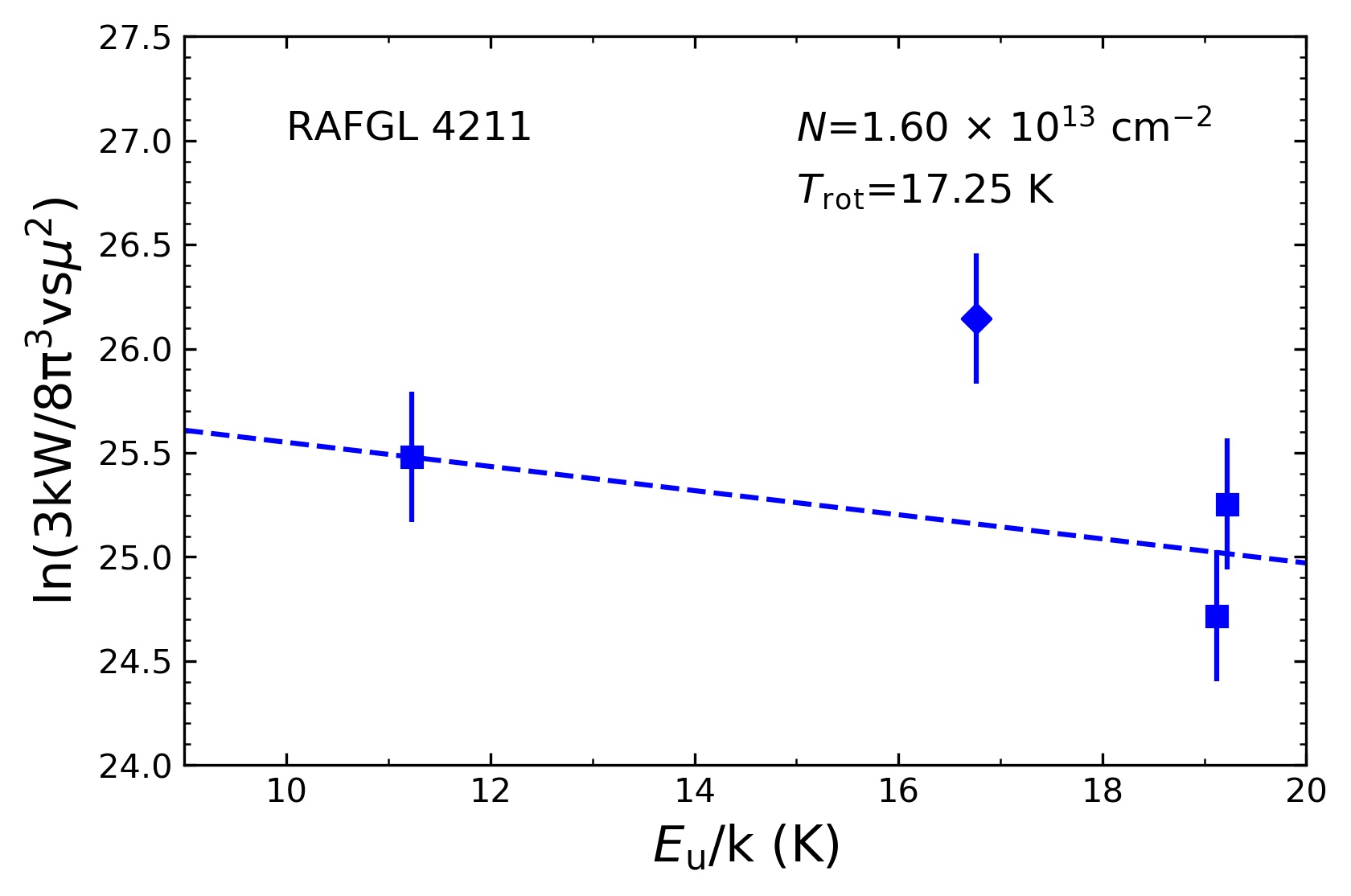}
\includegraphics[width=6cm]{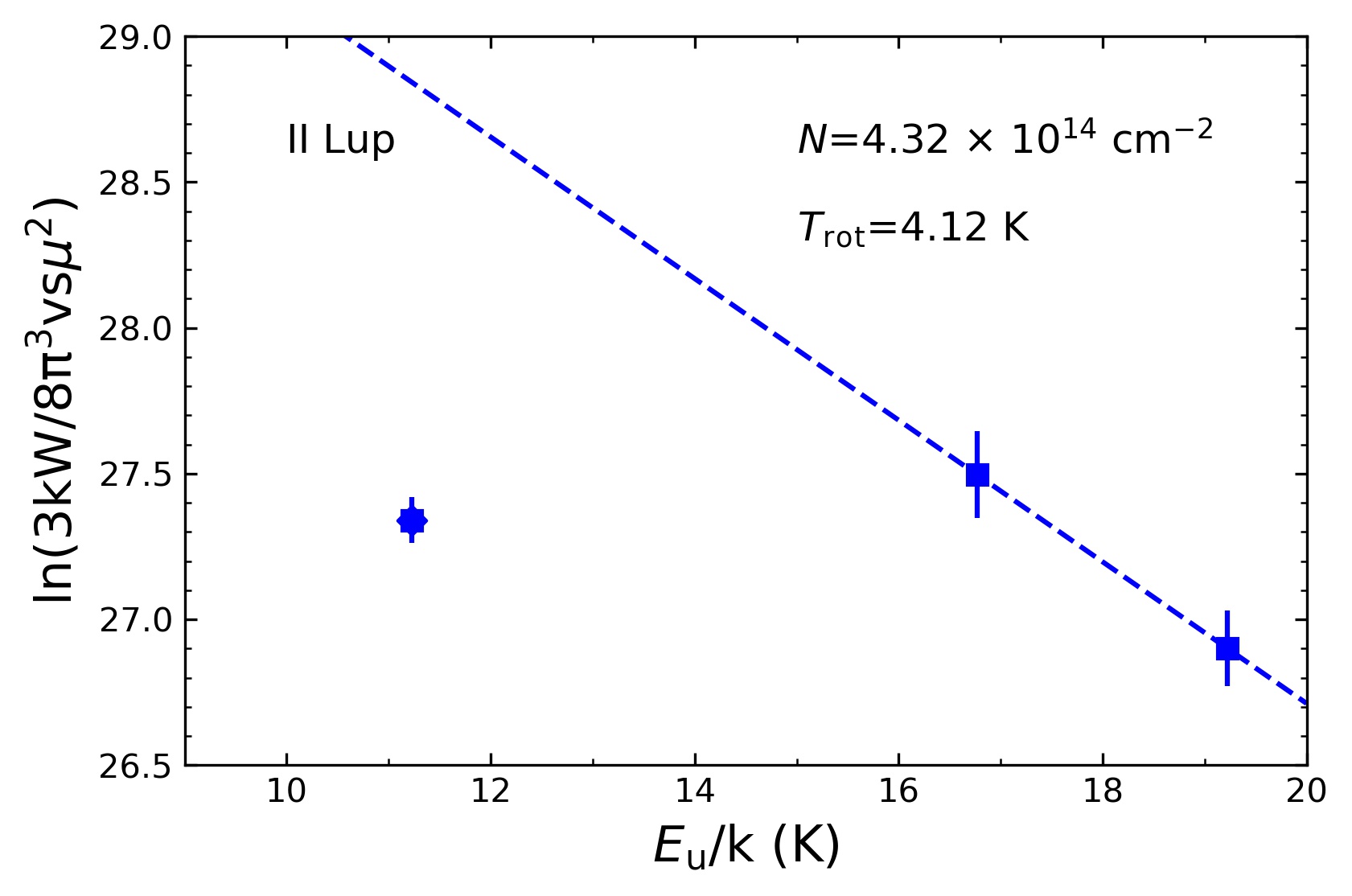}
\end{center}
\caption{Rotational diagram for the observed SiC$_2$ lines toward AI Vol, RAFGL 4211, and II Lup. 
The blue dotted lines represent the results from linear least squares fitting by taking into account the weights of errors. 
During the fitting procedures, 
the data for the SiC$_2$ transitions of $5_{0,\,5}-4_{0,\,4}$ (blue diamond) in RAFGL 4211 and $4_{0,\,4}-3_{0,\,3}$ (blue diamond) in II Lup were excluded.
The upper right corner in each panel shows the fitted column density and excitation temperature.} 

\label{fig:7}
\end{figure}

In the practical calculations, firstly, the excited temperature and column density of SiC$_2$ are fitted by using the rotation diagram method that described in equation \ref{equ:1},  and plotted in Figure \ref{fig:7}; secondly, 
the fractional abundance of SiC$_2$ is calculated by $f$ = $N/N_{\rm H_2}$. 
In AI Vol, we obtained $f$ (SiC$_2$) = 1.55 $\times$ $10^{-8}$,
which is an order of magnitude lower than the SiC$_2$ abundance in IRC+10216, which is around $10^{-7}$ \citep{agundez2012molecular, velilla2018circumstellar, cernicharo2010high, 2014MNRAS.445.3289F}. 
Table \ref{table:3} presents the fractional abundances, column densities, and excitation temperatures of SiC$_2$ for all three stars.  
We found that the fractional abundance of SiC$_2$ increases with the increasing wind density in the three observed C-rich AGB stars.

In the previous astrochemical models of C-rich CSEs of AGB stars, SiC$_2$ is always treated as a parent species with an initial abundance of $\thicksim$ $10^{-5}$ \citep{li2014photodissociation, 2018A&A...611A..29M, 2022MNRAS.510.1204V}. However, this study clearly shows that this species is a daughter species and therefore the chemistry of SiC$_2$ in the chemical models needs to be reinvestigated.  

\subsection{SiC$_2$ chemistry} \label{sec:part3.3}
Previous studies of the silicon chemistry in the CSEs of AGB stars mainly arise from those models constructed for the carbon-rich AGB star IRC+10216. The detailed discussion on the chemistry of the triangular molecule SiC$_2$ in the inner CSE can be found in \cite{Willacy98}, in the outer CSE in  \cite{1992PASJ...44..469T}, \cite{Gensheimer95}, and  \cite{Mackay99}, where SiC$_2$ is found to be mainly formed via ion-neutral reactions and destructed to SiC due to photodissociation induced by the photons from the interstellar medium. \cite{cernicharo2010high} introduced three key reactions to the formation of  SiC$_2$ (i.e., Si reacts with C$_2$H$_2$, Si reacts with C$_2$H, and Si$^+$ reacts with C$_2$H), and then can successfully explain the enhanced abundance of SiC$_2$ from observations. The chemistry of those cyclic molecules is complex. The study of c-SiC$_3$ by \citet{2019PNAS..11614471Y} suggested that the carbon-silicon molecules in the CSE may not only come from complex ion-molecule reactions or photodissociation of high-molecular-weight carbon-silicon molecules but also from bimolecular neutral-neutral reactions, and leading to the formation of naked carbon-silicon molecules via photochemical dehydrogenation. 
Since SiC$_2$ is likely the gas-phase precursor in forming SiC dust in carbon stars \citep{2018A&A...611A..29M}, the interaction between gas and dust, in addition to the non-spherical and clumpy structures of the envelopes of the stars, will need to be considered in the future chemical models.

\section{Conclusion} \label{sec:part4}

To explore whether SiC$_2$ is a parent or a daughter species, we have analyzed the ALMA observations of SiC$_2$ ($4_{0,\,4}-3_{0,\,3}$, $4_{2,\,3}-3_{2,\,2}$, $4_{2,\,2}-3_{2,\,1}$, and $5_{0,\,5}-4_{0,\,4}$) for the three carbon stars, and compared them with the SiC$_2$ results in the C-rich AGB star IRC +10216. The abundance of SiC$_2$ in the three selected stars is calculated by the rotational diagrams method. Our analysis revealed that the SiC$_2$ molecules in the CSEs of the carbon stars exhibited a hollow shell structure, with a distinct brightness distribution at different velocities, therefore we conclude that SiC$_2$ exists as a daughter molecule in the CSEs of these sources. Our results are in contrast to some previous reports which concluded that SiC$_2$ is a parent molecule in the envelopes of carbon stars. More sensitive observations may, of course, detect SiC$_2$ emission at low levels in the inner CSE. Our findings provide new insights into the chemical processes occurring in the CSEs of evolved stars and contribute to our understanding of the chemical evolution of ISM. Further detailed understanding of SiC$_2$ formation requires studying and comparing more C-rich AGB stars by combining the results of the single-dish and high-resolution interferometric observations.

\section*{Conflict of Interest Statement}

The authors declare that the research was conducted in the absence of any commercial or financial relationships that could be construed as a potential conflict of interest.
 
\section*{Author Contributions}

YF conducted the project, processed the ALMA archive data, and wrote the draft. XL initiated the project, guided the work, and revised the manuscript. TM was involved in interpreting the results and reviewing and revising the manuscripts. RS provided continuous manuscript revision. KW provided insights and feedback during the analysis process of the results. FX helped with the data analysis. DQ, SQ, XF, BJ, QC, GH, FL, and YZ provided comments and suggestions on the manuscript. JT, ZM, RM, JS, and JY were involved in analyzing the results and provided suggestions for the manuscript. All authors discussed and contributed to the final version of the manuscript.

\section*{Funding}

X. Li acknowledges support from the Xinjiang Tianchi project (2019). TJM is grateful to the Leverhulme Trust for the award of an Emeritus Fellowshiop. This work was also funded by the National Science Foundation of China (12173023), the China Manned Space Project (CMS-CSST-2021-A09), the National Science Foundation of China (11973013), and the National Key Research and Development Program of China (22022YFA1603102). National Natural Science Foundation of China (92261101) and the Innovation Capability Support Program of Shaanxi Province (2023-CX-TD-49). National Natural Science Foundation of China (11973075).
X. Fang, S. Qin, and Y. Zhang thank the Xinjiang Uygur Autonomous Region of China for support from the Tianchi Talent Program.

\section*{Acknowledgments}

We appreciate the invaluable suggestions of the two anonymous reviewers on this work, 
which has been significantly improved the quality of this paper. 
This paper makes use of the following ALMA data: ADS/JAO.ALMA 2013.1.00070.S. ALMA is a partnership of ESO (representing its member states), NSF (USA), and NINS (Japan), together with NRC (Canada), MOST and ASIAA (Taiwan), and KASI (Republic of Korea), in cooperation with the Republic of Chile. The Joint ALMA Observatory is operated by ESO, AUI/NRAO, and NAOJ. We are very grateful to Professor Lars-$\rm \AA$ke Nyman (the PI of the ALMA project that provided the data for this work) for his valuable comments and suggestions.  This research has made use of astropy\footnote{www.astropy.org}, a community-developed core Python package for astronomy, astropy, and matplotlib.



\section*{Data Availability Statement}

The datasets analyzed for this study can be found in the ALMA Science Archive \footnote{https://almascience.nrao.edu}.

\bibliographystyle{Frontiers-Harvard} 
\bibliography{SiC2}
\section*{Figure captions}

\begin{table}[h]
\renewcommand\arraystretch{1.5}
\small
 \caption{Source and observational parameters.}
 \label{table:1}
 \begin{tabular}{ccccccc}
  \hline
		Name    & IRAS&  Mass loss rate                &  Distance    &  $V\rm_{LSR}$    &  $ V\rm_{exp}$      &  $ \theta\rm_{beam}$   \\
		        &     & ($M\rm_{\odot}$ $\rm yr^{-1}$) &  (pc)        &  (km s$^{-1}$)   &  (km s$^{-1}$)      &  ($^{\prime \prime}$)   \\
  \hline
		AI Vol      & IRAS 07454-7112  & $4.9(-6)^{(2)}$  & $710^{(2)}$   & $-39.0^{(2)}$    & $12.0^{(2)}$   &  1.51 $\times$ 0.99$^{(1)}$   \\
		RAFGL 4211  & IRAS 15082-4808  & $1.0(-5)^{(3)}$  & $850^{(4)}$   & $-3.0^{(5)}$     & $19.5^{(3)}$   &  1.00 $\times$ 1.00$^{(1)}$   \\
		II Lup      & IRAS 15194-5115  & $1.7(-5)^{(2)}$  & $500^{(2)}$   & $-15.5^{(2)}$    & $21.5^{(2)}$   &  0.75 $\times$ 0.45$^{(1)}$   \\
  \hline
 \end{tabular}\\
{\bf Note.} $a(b)=a\times10^b$. Mass loss, distance, local standard of rest ($V\rm_{LSR}$) and expansion ($V\rm_{exp}$) velocities from
the literature, and synthesized beam, $\rm \theta _{beam}$. References: $^{(1)}$ This work. $^{(2)}$\citet{danilovich2018sulphur}. $^{(3)}$\citet{woods2003molecular}. $^{(4)}$\citet{groenewegen2002millimetre}. $^{(5)}$\citet{smith2015molecular}.
\end{table}

\begin{table}[h]
\begin{center}
\caption[]{Spectral line parameters of different transitions in SiC$_2$ obtained via shell fitting using GILDAS.}
\setlength{\tabcolsep}{2.5pt}
\LARGE
\renewcommand\arraystretch{2}
 \label{table:2}
 \resizebox{\textwidth}{!}{
 \begin{tabular}{cccccccccccccccc}
  \hline\noalign{\smallskip}
    & & \multicolumn{4}{c}{AI Vol} & & \multicolumn{4}{c}{RAFGL 4211}  & &   \multicolumn{4}{c}{II Lup}   \\
      \cline{3-6}        \cline{8-11}    \cline{13-16}
  Transition & Frequency & rms & $\int T\rm_{R}d\upsilon$ & $T\rm_{peak}$ & $S/N$ &&  rms   & $\int T\rm_{R}d\upsilon$ & $T\rm_{peak}$ & $S/N$ &&  rms  & $\int T\rm_{R}d\upsilon$ & $T\rm_{peak}$  & $S/N$ \\
             &   (GHz)   & (mK) & (K km s$^{-1}$)     &    (K)  &    &&  (mK)  & (K km s$^{-1}$)           &    (K) &    && (mK)  &  (K km s$^{-1}$)         &    (K)  &   \\
   \hline
$J_{Ka,\,Kc}$ = $4_{0,\,4}-3_{0,\,3}$ &  93.06363900 & 6.70 &  1.26 ($\pm$ 0.21)   &  0.14  & 20.45 &&  16.40 &  1.48 ($\pm$ 0.52)             & 0.14  & 8.60 &&  21.49 &     9.51 ($\pm$ 0.75)            &  0.54 & 25.11  \\
$J_{Ka,\,Kc}$ = $4_{2,\,3}-3_{2,\,2}$ &  94.24539300 & 2.73 &  0.29 ($\pm$ 0.09)   &  0.03  & 10.99 &&  10.70 &  0.53 ($\pm$ 0.34)             & 0.06  & 5.51 &&  25.13 &     6.82 ($\pm$ 0.88)            &  0.37 & 14.82  \\
$J_{Ka,\,Kc}$ = $4_{2,\,2}-3_{2,\,1}$ &  95.57938100 & 2.55 &  0.32 ($\pm$ 0.08)   &  0.04  & 13.73 &&  16.30 &  0.91 ($\pm$ 0.52)             & 0.10  & 6.13 &&  47.30 &     $-$                     &  $-$     & $-$  \\
$J_{Ka,\,Kc}$ = $5_{0,\,5}-4_{0,\,4}$ & 115.38238880 & 6.64 &  0.48 ($\pm$ 0.21)   &  0.04  & 5.27  &&  40.60 &  4.45 ($\pm$ 1.30)             & 0.33  & 8.08 &&  72.80 &     17.20 ($\pm$ 2.55)           &  0.80 & 11.03  \\
\noalign{\smallskip}\hline
\end{tabular}}
\end{center}
{\bf Note}. The uncertainties are indicated in parentheses.
The S/N is calculated by $T\rm_{peak}$/rms.

\end{table}

\begin{table}[h]
\begin{center}
\caption[]{The deduced excitation temperature ($T\rm_{ex}$), column density ($N$), fractional abundance ($f$) of SiC$_2$ relative to H$_2$ in the target sources by the rotational diagram method, along with the previous observational results.
}
\setlength{\tabcolsep}{2.5pt}
\small
\renewcommand\arraystretch{2}
 \label{table:3}
  \resizebox{\textwidth}{!}{
 \begin{tabular}{ccccccccc}
  \hline\noalign{\smallskip}
    & \multicolumn{3}{c}{This work} &  & \multicolumn{4}{c}{Other observations}     \\
      \cline{2-4}        \cline{6-9}
  Source        & $T\rm_{ex}$ &  $N$$\times$ 10$^{13}$         &       $f$ $\times$ 10$^{-8}$                  & &  $ T\rm_{ex}$     &    $N$$\times$ 10$^{13}$      &  $f$ $\times$ 10$^{-7}$        & $\dot{M}$ $/$ $V$$\rm_{exp}$$\times$ 10$^{-7}$      \\
                & (K)         &  (cm$^{-2}$)                    &                            & &   (K)             &    (cm$^{-2}$)             &            &
  ($M\rm_{\odot}$ $\rm yr^{-1}$ km$^{-1}$ s)    \\
   \hline
  AI Vol     & 6.66 ($\pm$ 1.33)   &  0.90 ($\pm$ 0.49)   & 1.64  & &   14.00 ($\pm$ 16.00)$^{(b)}$    & 0.40 ($\pm$ 0.50)$^{(b)}$ & 2.30$^{(a)}$  & 4.08$^{(c)}$\\
  RAFGL 4211 & 17.25 ($\pm$ 13.07) &  1.60 ($\pm$ 2.10)   & 4.62  & &   35.00 ($\pm$ 51.00)$^{(b)}$    & 4.00 ($\pm$ 3.00)$^{(b)}$ & 4.90$^{(a)}$  & 5.13$^{(c)}$\\
  II Lup     & 4.12   &  43.22 & 47.46 & &   19.00 ($\pm$ 11.00)$^{(b)}$    & 4.00 ($\pm$ 2.00)$^{(b)}$ & 12.00$^{(a)}$ & 7.91$^{(a,\,c)}$\\
\noalign{\smallskip}\hline
\end{tabular}}
\end{center}
{\bf Note}. 
$\dot{M}$ $/$ $V\rm_{exp}$: the CSEs wind density. 
The uncertainties are listed in parentheses, 
which are obtained by considering the 1$\sigma$ signal via the error propagation. 
The fitting of II Lup was performed using only two data points, which resulted in a lower precision and the absence of error estimation.
References: $^{(a)}$ \citet{woods2003molecular};$^{(b)}$ \citet{smith2015molecular};$^{(c)}$ \citet{danilovich2018sulphur}.
\end{table}

\begin{table}[h]
\centering
\renewcommand\arraystretch{1.5}
\small
 \caption{
 Upper energy, $E$$\rm_u$, and $S\mu^2$ of molecular SiC$_2$ in the observed transitions,
extracted from the ``Splatalogue'' database.
 }
 \label{table:4}
 \begin{tabular}{ccc}
  \hline
	Transition     &   $E$$\rm_u$      &  $S\mu^2$     \\
                                     &    (K)            &   (Debye$^2$)     \\
  \hline
$J_{Ka,\,Kc}$ = $4_{0,\,4}-3_{0,\,3}$  &  11.23 & 22.82   \\
$J_{Ka,\,Kc}$ = $4_{2,\,3}-3_{2,\,2}$  &  19.12 & 17.18  \\
$J_{Ka,\,Kc}$ = $4_{2,\,2}-3_{2,\,1}$  &  19.22 & 17.18  \\
$J_{Ka,\,Kc}$ = $5_{0,\,5}-4_{0,\,4}$  &  16.77 & 28.46  \\
  \hline
 \end{tabular}\\
\end{table}

\appendix
\section{Supplementary Figures}




\renewcommand\thefigure{\Alph{section}\arabic{figure}}
\setcounter{figure}{0}
\begin{figure}[h!]
\begin{center}
\includegraphics[width=15cm]{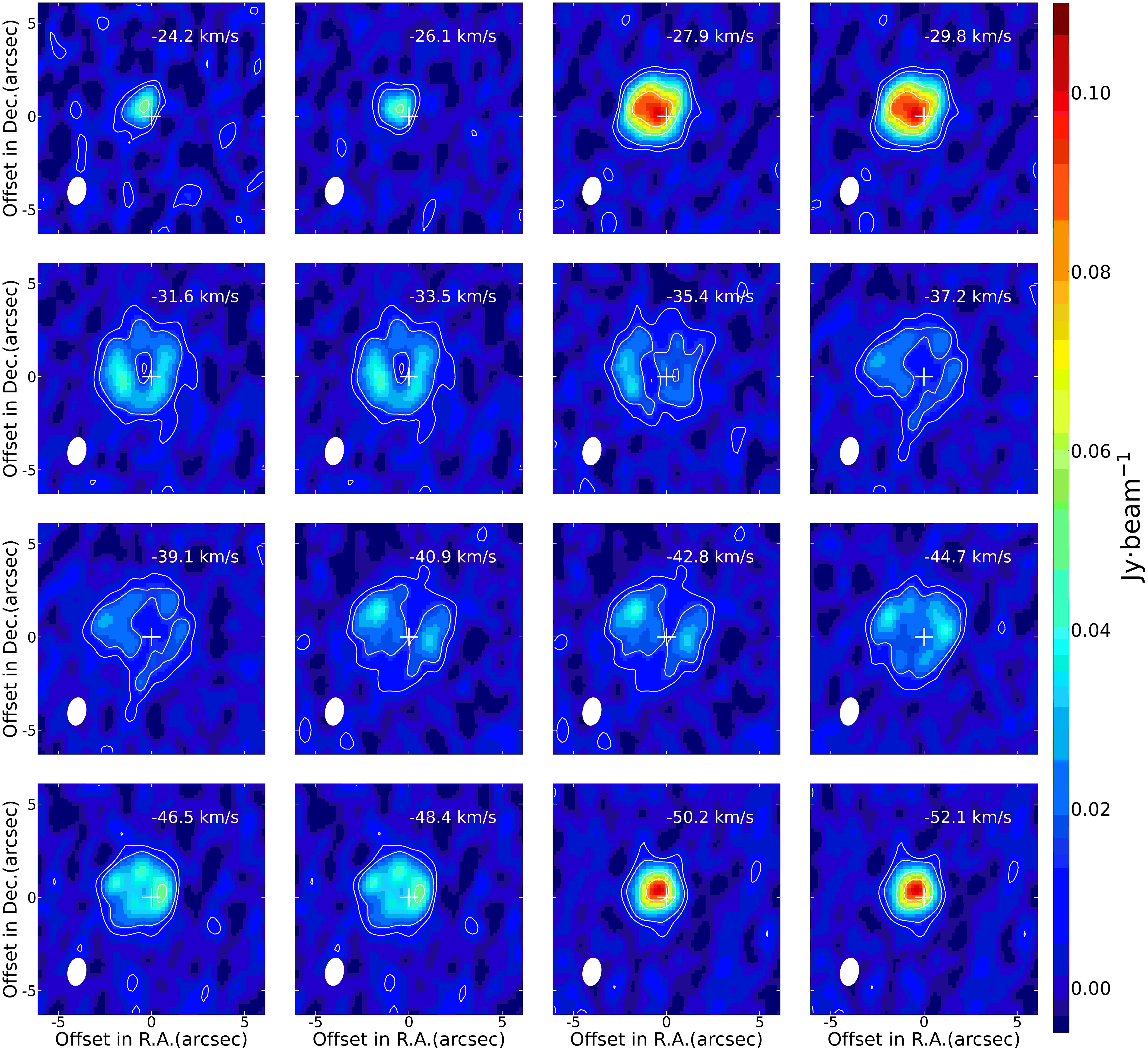}
\end{center}
\caption{Channel maps of SiC$_2$ $(4_{2,\,3} - 3_{2,\,2})$ towards AI Vol. The white ellipse in the lower left corner of each channel diagram is the beam. The beam size of observation is 1.51 $\times$ 0.99 $^{\prime \prime}$ and PA 9.82$^{\circ}$. The white contours delineate the flux levels of the SiC$_2$, indicating increments of 5, 10, 30, and 50 times the rms noise (1$\sigma$ = 1.6 mJy$\cdot$beam$^{-1}$).}
\label{subfig:1}
\end{figure}

\begin{figure}[h!]
\begin{center}
\includegraphics[width=15cm]{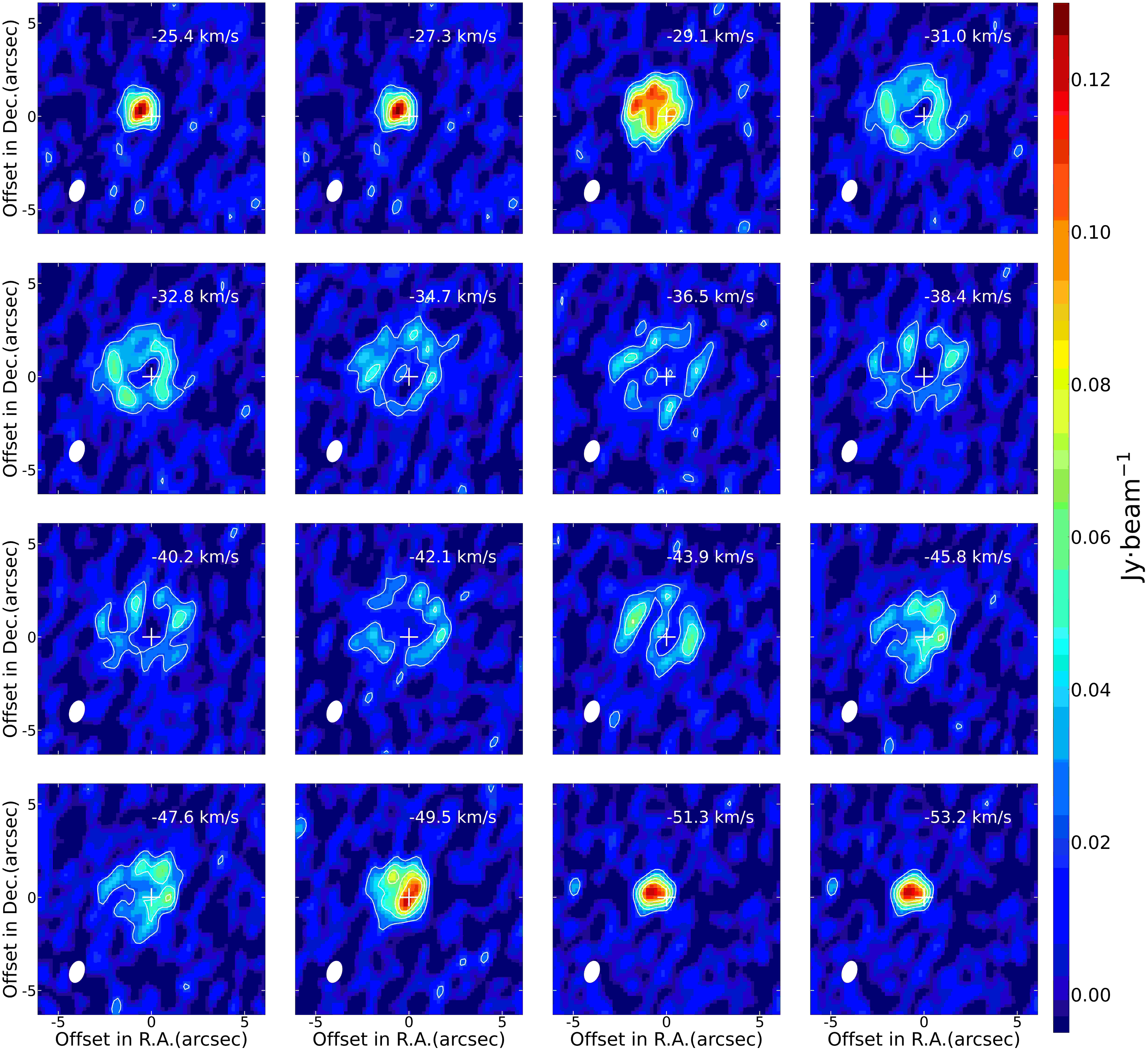}
\end{center}
\caption{Channel maps of SiC$_2$ $(5_{0,\,5} - 4_{0,\,4})$ towards AI Vol. The white ellipse in the lower left corner of each channel diagram is the beam. The beam size of observation is 1.20 $\times$ 0.79 $^{\prime \prime}$ and PA 18.86$^{\circ}$. The white contours delineate the flux levels of the SiC$_2$, indicating increments of 5, 10, 15, and 20 times the rms noise (1$\sigma$ = 4.3 mJy$\cdot$beam$^{-1}$).}
\label{subfig:2}
\end{figure}

\begin{figure}[h!]
\begin{center}
\includegraphics[width=15cm]{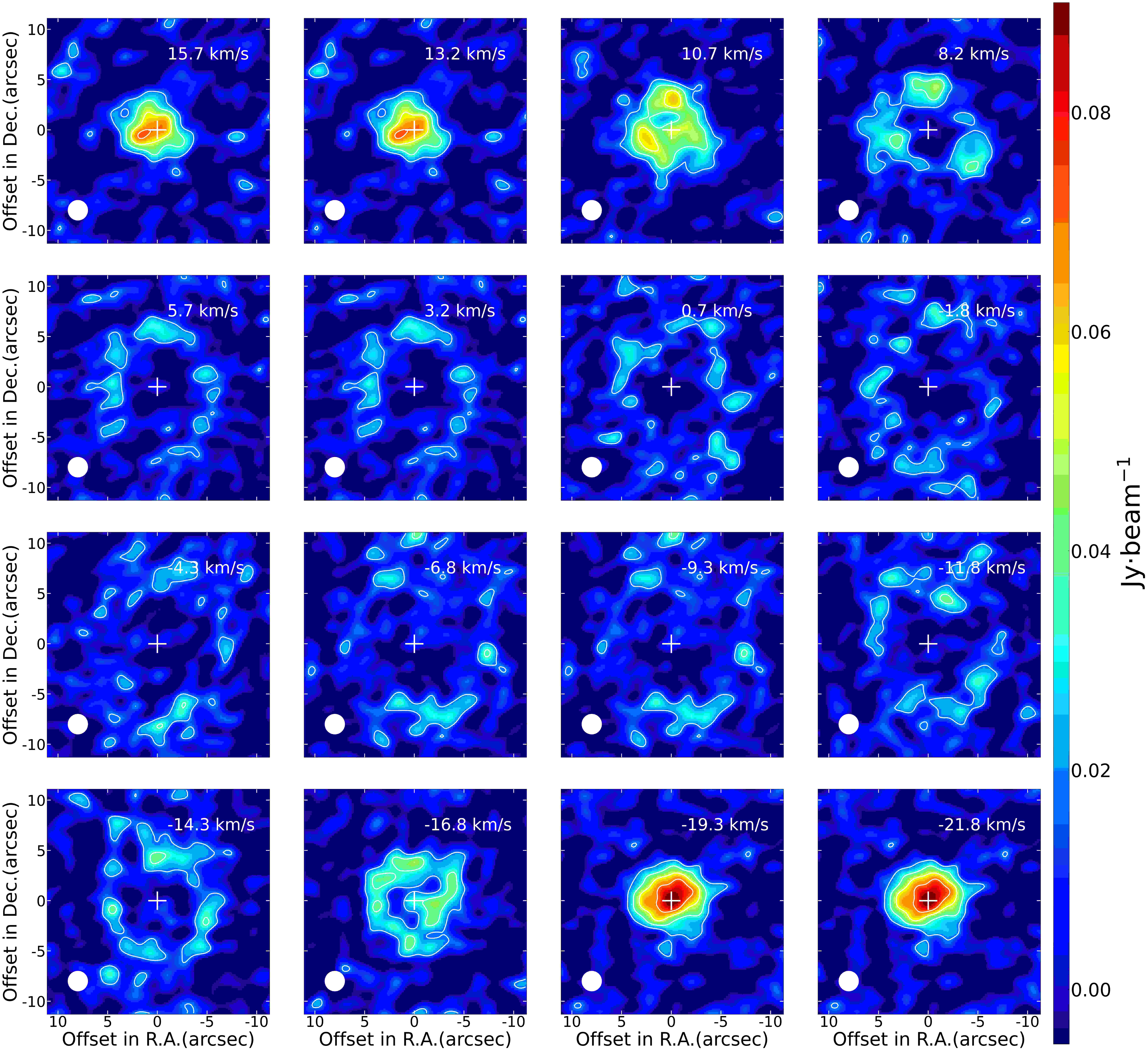}
\end{center}
\caption{Channel maps of SiC$_2$ $(4_{0,\,4} - 3_{0,\,3})$ towards RAFGL 4211. The white circle in the lower left corner of each channel diagram is the beam. The beam size of observation is 1.00 $\times$ 1.00 $^{\prime \prime} $ and PA 0$^{\circ}$. 
The white contours delineate the flux levels at 5, 10, 15, and 20 times the rms noise (1$\sigma$ = 3.5 mJy$\cdot$beam$^{-1}$).}
\label{subfig:3}
\end{figure}

\begin{figure}[h!]
\begin{center}
\includegraphics[width=15cm]{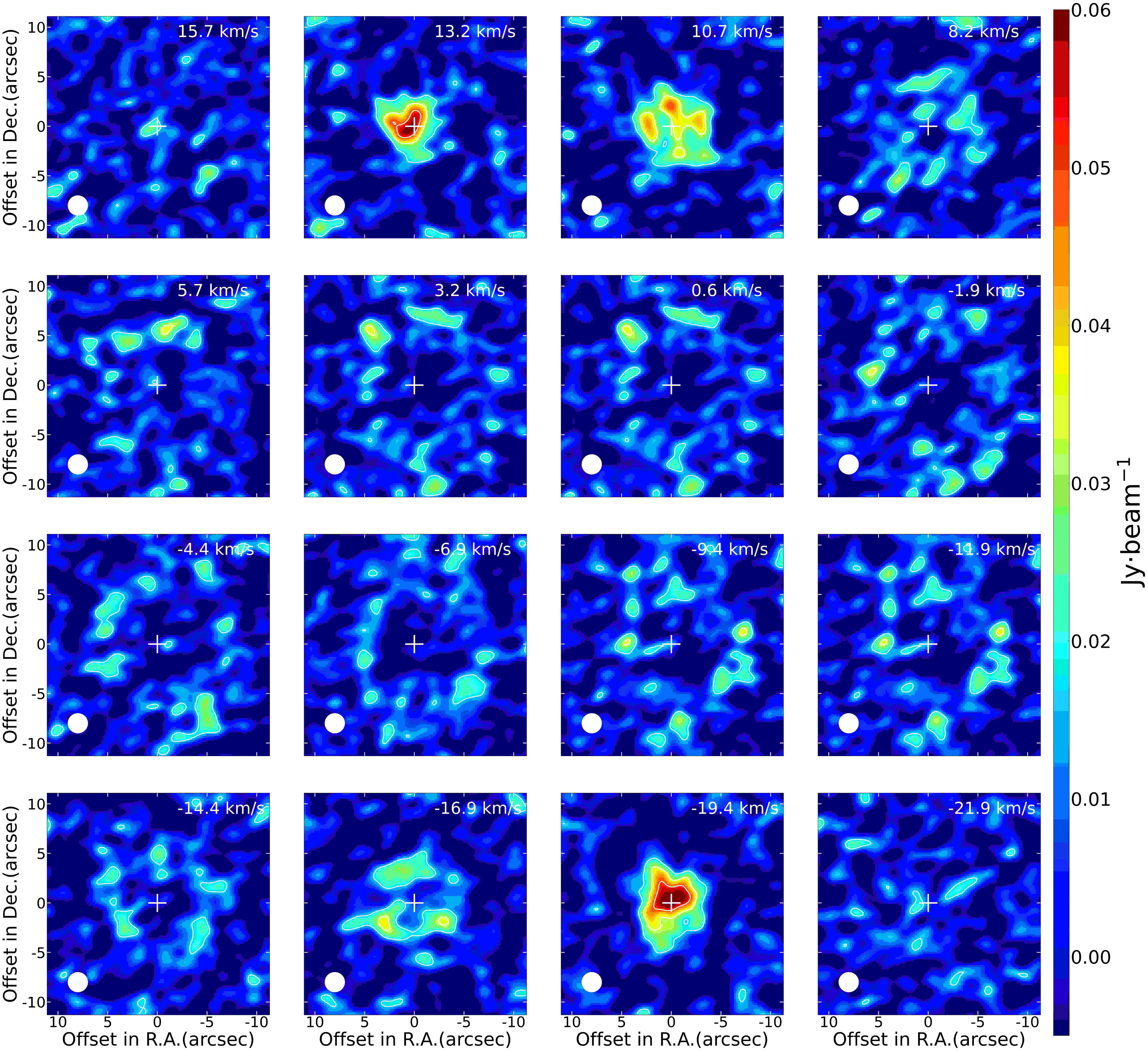}
\end{center}
\caption{Channel maps of SiC$_2$ $(4_{2,\,3} - 3_{2,\,2})$ towards RAFGL 4211. The white circle in the lower left corner of each channel diagram is the beam. The beam size of observation is 1.00 $\times$ 1.00 $^{\prime \prime}$ and PA 0$^{\circ}$. The white contours delineate the flux levels at 5, 10, and 15 times the rms noise (1$\sigma$ = 3.3 mJy$\cdot$beam$^{-1}$).}
\label{subfig:4}
\end{figure}

\begin{figure}[h!]
\begin{center}
\includegraphics[width=15cm]{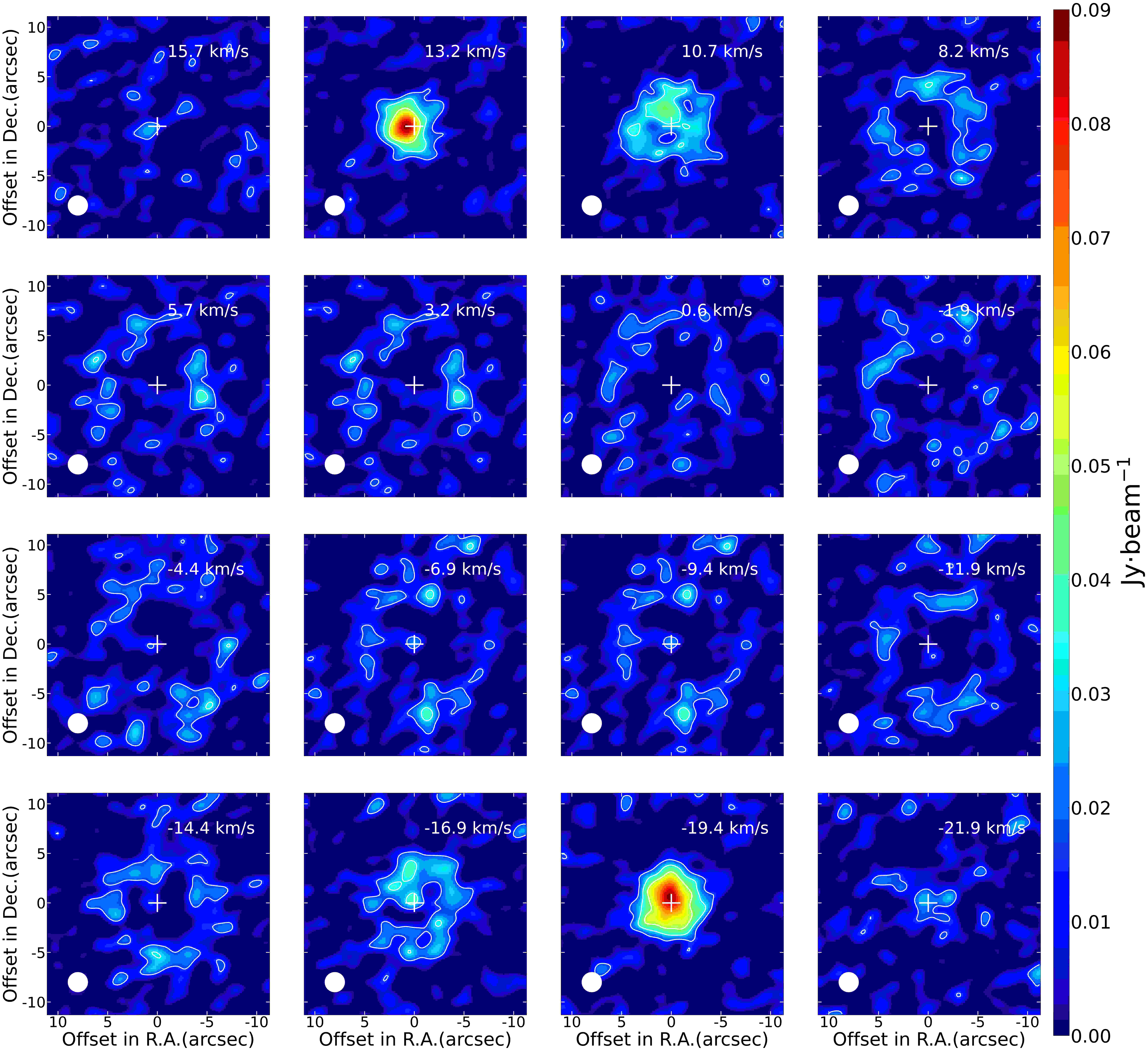}
\end{center}
\caption{Channel maps of SiC$_2$ $(4_{2,\,2} - 3_{2,\,1})$ towards RAFGL 4211. The white circle in the lower left corner of each channel diagram is the beam. The beam size of observation is 1.00 $\times$ 1.00 $^{\prime \prime}$ and 0$^{\circ}$. The white contours delineate the flux levels  noise (1$\sigma$ = 3.2 mJy$\cdot$beam$^{-1}$).}
\label{subfig:5}
\end{figure}

\begin{figure}[h!]
\begin{center}
\includegraphics[width=15cm]{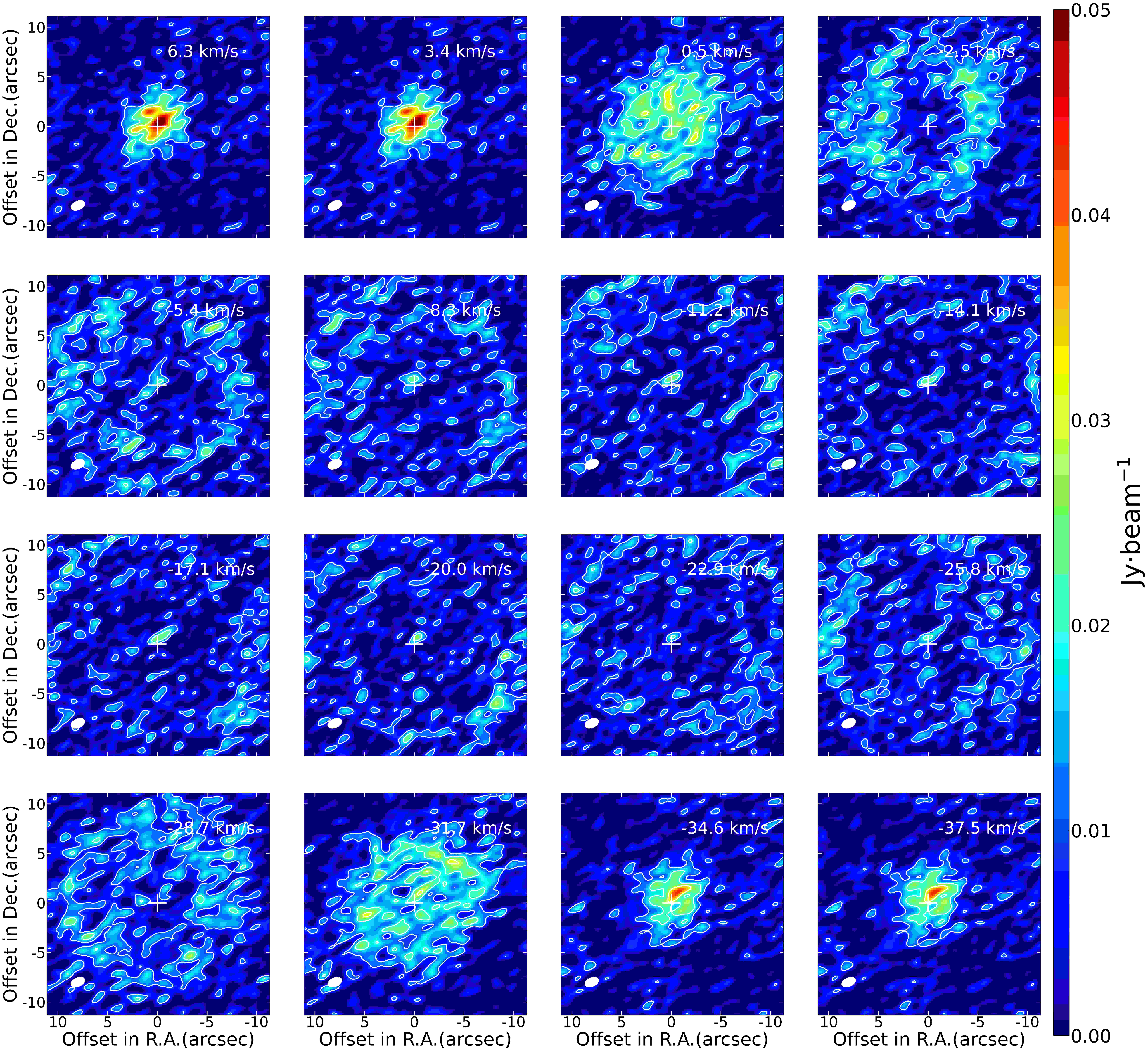}
\end{center}
\caption{Channel maps of SiC$_2$ $(4_{0,\,4} - 3_{0,\,3})$ towards II Lup. The white ellipse in the lower left corner of each channel diagram is the beam. The beam size of observation is 0.75 $\times$ 0.45 $^{\prime \prime}$ and PA 65.6$^{\circ}$. The white contours delineate the flux levels at 5, 10, and 15 times the rms noise (1$\sigma$ = 2.0 mJy$\cdot$beam$^{-1}$).}
\label{subfig:6}
\end{figure}

\begin{figure}[h!]
\begin{center}
\includegraphics[width=15cm]{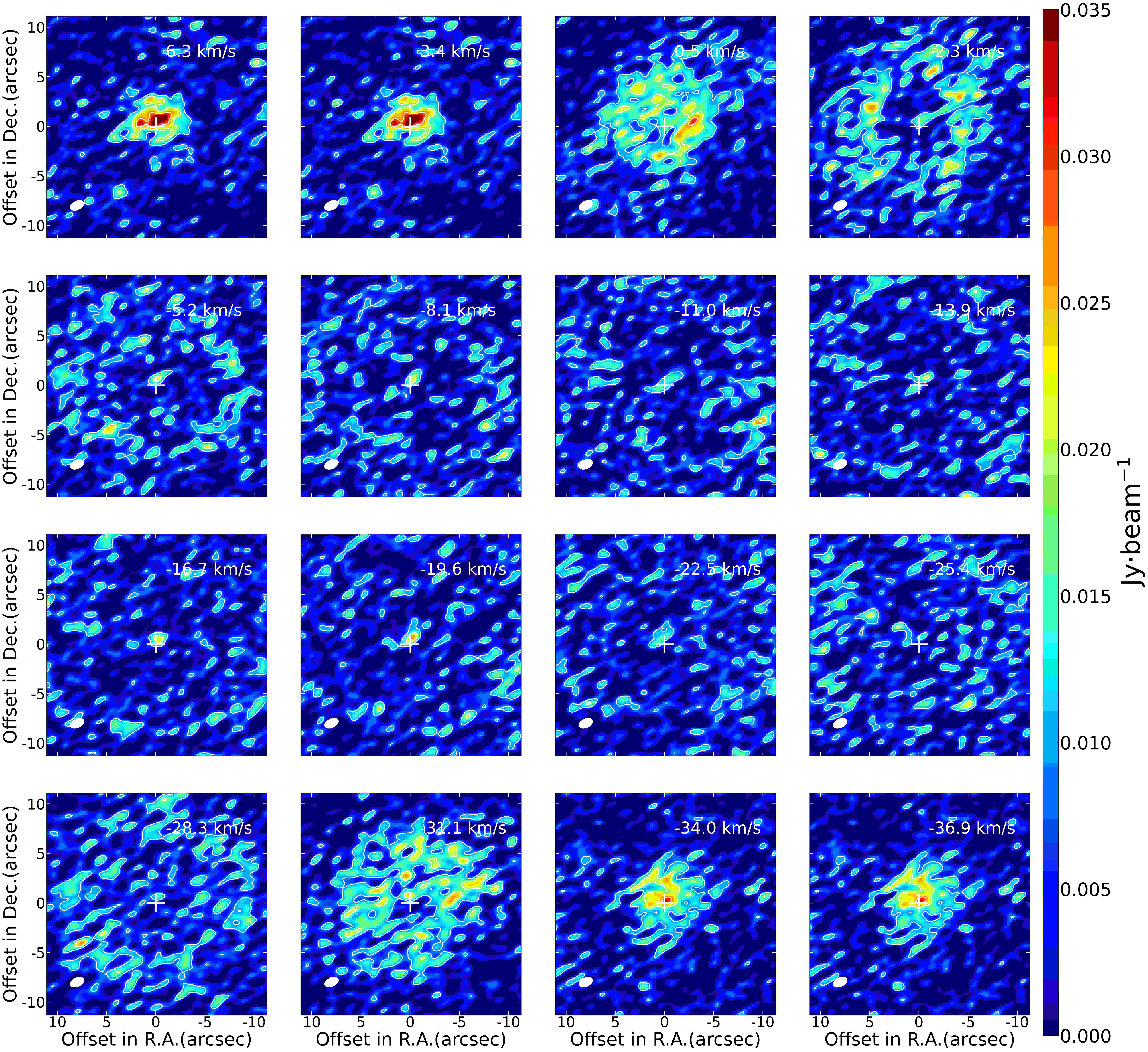}
\end{center}
\caption{Channel maps of SiC$_2$ $(4_{2,\,3} - 3_{2,\,2})$ towards II Lup. The white ellipse in the lower left corner of each channel diagram is the beam. The beam size of observation is 0.75 $\times$ 0.45 $^{\prime \prime}$ and 65.6$^{\circ}$. The white contours delineate the flux levels at 5, 10, and 15 times the rms noise (1$\sigma$ = 1.8 mJy$\cdot$beam$^{-1}$).}
\label{subfig:7}
\end{figure}

\begin{figure}[h!]
\begin{center}
\includegraphics[width=16cm]{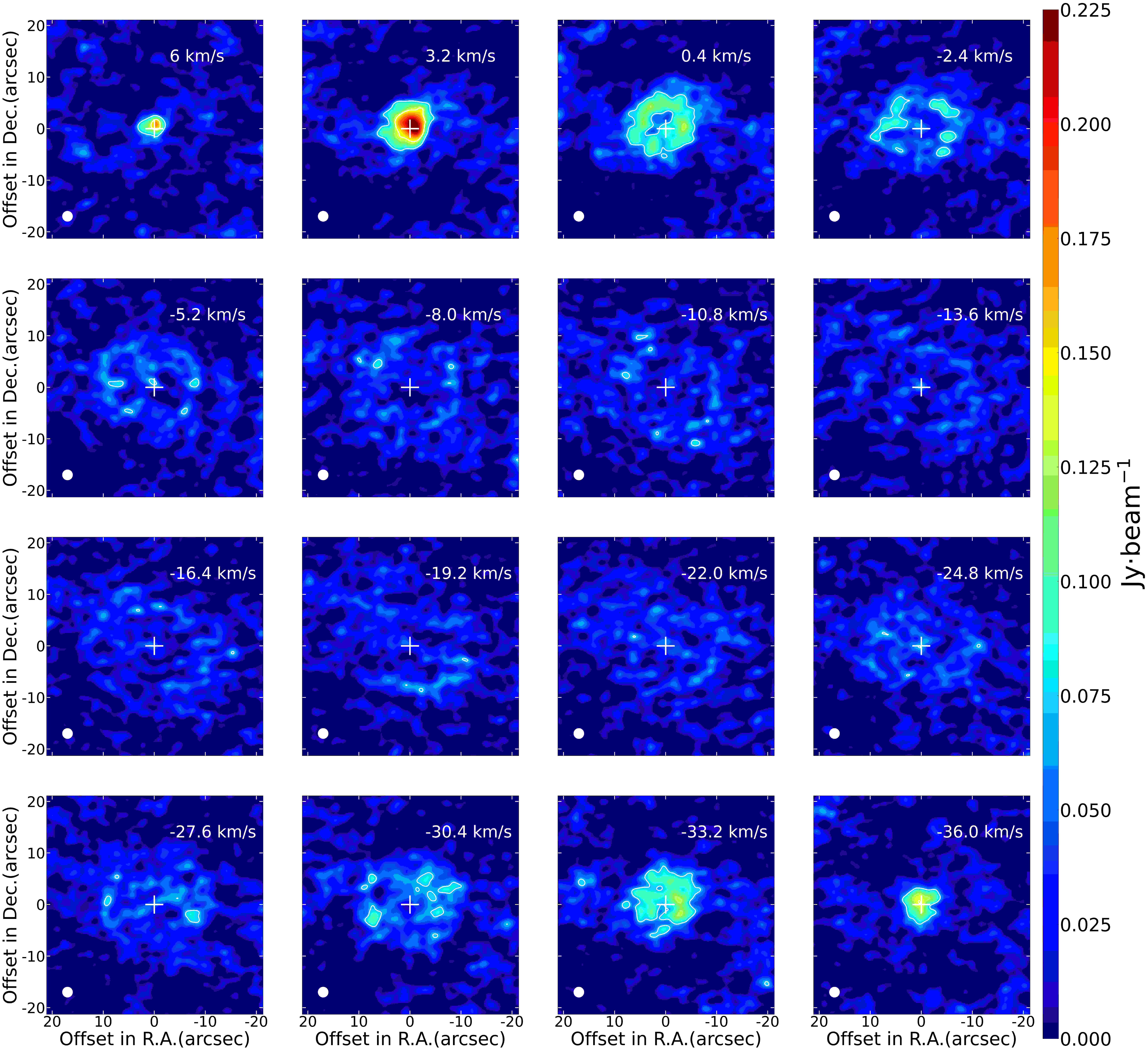}
\end{center}
\caption{Channel maps of SiC$_2$ $(5_{0,\,5} - 4_{0,\,4})$ towards II Lup. The white circle in the lower left corner of each channel diagram is the beam. The beam size of observation is 1 $\times$ 1 $^{\prime \prime}$ and PA 0$^{\circ}$. The white contours delineate the flux levels at 5, 10, and 20 times the rms noise (1$\sigma$ = 13.8 mJy$\cdot$beam$^{-1}$).}
\label{subfig:8}
\end{figure}

\begin{figure}[h!]
\begin{center}
\includegraphics[width=18cm]{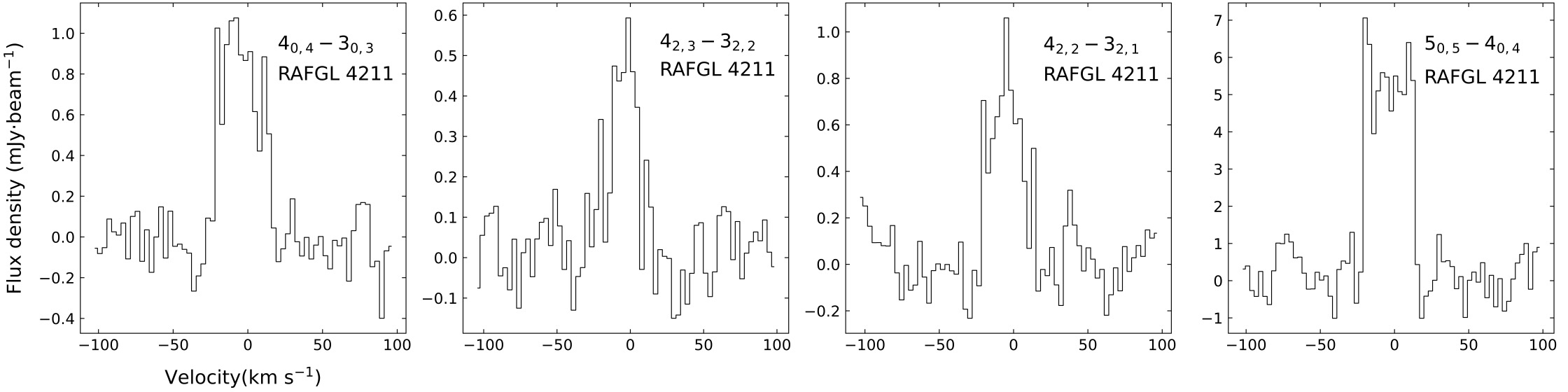}
\quad
\includegraphics[width=18cm]{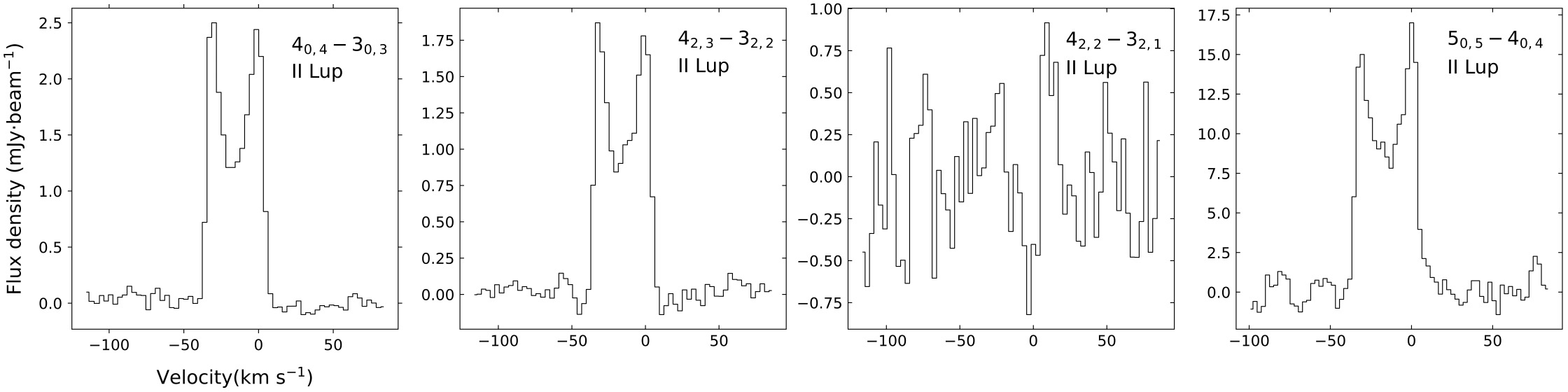}
\end{center}
\caption{The spectral lines of SiC$_2$ toward RAFGL 4211 and II Lup. The top four panels are RAFGL 4211. The bottom four panels are II Lup. The four transition spectra of these two sources were obtained by integrating with a 100 $\times$ 100 pixel range centered on the star.}
\label{subfig:9}
\end{figure}


\end{document}